\documentclass[journal,12pt,onecolumn,draftclsnofoot]{IEEEtran}
\usepackage{setspace} 
\usepackage{cite}
\usepackage{microtype}
\usepackage{cprotect}
\usepackage{float}
\usepackage{graphicx}
\usepackage{capt-of}
\usepackage{subcaption}
\usepackage{booktabs} 

\usepackage{makecell}
\usepackage{amsmath}
\usepackage{authblk}
\usepackage{wrapfig}
\usepackage{algorithm}
\usepackage{forloop}
\usepackage{color}
\usepackage{cite}
\usepackage{algpseudocode}
\usepackage{indentfirst}
\usepackage{multirow}
\usepackage{enumitem}
\usepackage{fancyvrb}
\usepackage{fancyhdr}

\newtheorem{assumption}{Assumption}

\newcommand{\LinearSTHawkes}{\Verb|LinearSTHawkes|}
\newcommand{\NonLinearSTHawkes}{\Verb|NonLinearSTHawkes|}

\newcommand{\thetabeta}{\theta[\beta]}

\usepackage{amsfonts}

\newcommand{\ERAPS}{\Verb|ERAPS|}
\newcommand{\SRAPS}{\Verb|SRAPS|}

\newcommand{\rev}[1]{{\color{black}#1}}
\newcommand{\revtwo}[1]{{\color{black}#1}}

\newcommand{\R}{\mathbb{R}}
\newcommand{\PP}{\mathbb{P}}
\newcommand{\A}{\mathcal{A}}
\newcommand{\tildeF}[1]{\tilde{F}(#1)}
\newcommand{\hatF}[1]{\hat{F}(#1)}

\newcommand{\kreg}{k_{reg}}
\usepackage{authblk}
\usepackage{hyperref}
\usepackage[hang]{footmisc} 
\title{Spatio-Temporal Wildfire Prediction using Multi-Modal Data}

\author[1]{Chen Xu} 
\author[1]{Yao Xie\footnote{Corresponding author: Yao Xie (yao.xie@isye.gatech.edu)}} 
\author[2]{\\ Daniel A. Zuniga Vazquez}
\author[2]{Rui Yao}
\author[2]{Feng Qiu}
\affil[1]{{\small H. Milton Stewart School of Industrial and Systems Engineering, Georgia Institute of Technology.}}
\affil[2]{{\small Energy Systems Division, Argonne National Laboratory}}
\begin{document}

\maketitle
\vspace{-0.65in}
\begin{abstract}


Due to severe societal and environmental impacts, wildfire prediction using multi-modal sensing data has become a highly sought-after data-analytical tool by various stakeholders (such as state governments and power utility companies) to achieve a more informed understanding of wildfire activities and plan preventive measures. A desirable algorithm should precisely predict fire risk and magnitude for a location in real time. In this paper, we develop a flexible spatio-temporal wildfire prediction framework using multi-modal time series data. We first predict the wildfire risk (the chance of a wildfire event) in real-time, considering the historical events using discrete mutually exciting point process models. Then we further develop a wildfire magnitude prediction set method based on the flexible distribution-free time-series conformal prediction \rev{(CP)} approach. Theoretically, we prove a risk model parameter recovery guarantee, as well as coverage and set size guarantees for the \rev{CP} sets. Through extensive real-data experiments with wildfire data in California, we demonstrate the effectiveness of our methods, as well as their flexibility and scalability in large regions. 
\end{abstract}

\section{Introduction}

In recent years, widespread large-scale wildfire cause severe consequences, including direct property damage and economic losses, community evacuation, and fatalities, as well as impacts on nature such as higher CO$_2$ emissions \cite{Andreae2001EmissionOT}. To monitor and prevent severe consequence caused by large-scale wildfire, an imperative challenge was brought up: how to utilize multi-modal data collected through various sensing technologies, so as to precisely predict wildfire risk and magnitude for a local region and monitor the predictions in real-time. 

Wild fire risk prediction is particularly important for power utility companies to enhance their capability in making precise location-wise wildfire risk predictions. 
To prevent damage and economic losses, the utility companies also perform schedule utility shutdown for high wild-fire risk regions \cite{caWildfireWildfire}. Despite such urgent and essential need, utility companies often only leverage simple models/metrics for risks assessment, such as the burning index (BI) \cite{noaaFireWeather} and the fire load index \cite{nwcgNFDRSSystem}, which are static metrics that do not take into account the contribution from historical wildfire incidents and auxiliary environmental information. Imprecise wildfire risk prediction is causing sub-optimal power operator actions (such as unnecessary shut-down) that significantly disrupt reliable power delivery to customers.


Meanwhile, thanks to the development of sensing technology, there have been abundant multi-modal data collected through a variety of sensing mechanisms to gather wildfire information \cite{govDetectingWildfire}, which provides the unique opportunity for using sensing to perform precise location-wise real-time wildfire prediction. Common approaches to identify wildfire incidents include reports from human observers, wireless sensing \cite{Srinivasan2007ASO}, and infrared technology. Additional environmental information (e.g., weather and environmental conditions) has been integrated with each record, thus providing excellent opportunities for subsequent statistical analyses. As a result, each wildfire record is multi-modal: we know not only when and where it occurred but also its magnitude, the condition of the surrounding (e.g., infrastructure type), current weather information, and so on. Nevertheless, most existing wildfire modeling approaches \cite{Lee2002InformationSI,Wotton2007InterpretingAU,Jain2020ARO,Jaafari2019HybridAI} have not been designed to utilize such abundant multi-modal data.

In this paper, we present a framework for predicting wildfire risk and magnitude using multi-modal sensing data, based on a mutually exciting point process model and time series conformal prediction sets. Our model can capture the complex \rev{spatial-temporal} dependence of the multi-modal data through mutually exciting point processes, which is a natural framework for real-time prediction, since the conditional probability can be used to \rev{capture} fire risk given the past observations. In addition, we present a fire magnitude prediction algorithm through time-series \rev{CP} sets. Theoretically, we first prove model parameter recovery guarantees of the point process model for risk prediction. We then present coverage guarantees of fire magnitude prediction sets. Through extensive real-data experiments, we verify our models' competitive performances against other baseline methods regarding the precision of wildfire risk prediction.

Our prediction framework has the following features: (i) Predicting the wildfire risk --- the chance of \textit{binary} fire event (no fire versus fire) at a given locations and times, given historical observations and available multi-modal data (which can be treated as marks of the point processes), using a flexible marked spatio-temporal Hawkes process model \cite{Hawkes1971SpectraOS}. Specifically, we model the {\it mutual exciting property} in that historical and neighboring occurrences likely affect the occurrence likelihood, where certain occurrences may increase the chance while others inhibit the chance. The model parameters are efficiently estimated using an alternating optimization approach, in contrast to the more expensive expectation-maximization method \cite{reinhart2017review}. (ii) Exploiting interdependence among different geographic regions and the mutually exciting point process model is highly interpretable. (iii) Predicting fire magnitude using time-series \rev{CP} set, which can guarantee to contain true fire magnitude with a user specified high probability.

The rest of the paper is organized as follows. Section \ref{sec_data} describes background on sensing and the wildfire dataset. Section \ref{sec:mtd} contains our proposed methods. In particular, Section \ref{sec:mtd_hawkes} introduces proposed spatio-temporal Hawkes process models, which either linearly (i.e., \LinearSTHawkes) or nonlinearly (i.e., \NonLinearSTHawkes) quantify feature contributions to fire hazards. Section \ref{sec:mtd_hawkes_est} describes the objective function, the estimation procedure, and how to yield binary predictions based on predicted risks. Section \ref{sec:conformal} describes the \rev{CP} sets for wildfire magnitude prediction. Section \ref{theory} has two parts. We first present the theoretical analyses regarding the accuracy of fire risk prediction as a result of model recovery guarantee in Section \ref{theory_param_recovery}. Section \ref{theory_conformal} then verifies coverage guarantee of the prediction sets, whose size also converge to the true fire sizes asymptotically. Section \ref{sec:exp} first validates the proposed model on a small-scale real-data experiment, where Section \ref{sec:exp_Hawkes_carved} compares \LinearSTHawkes \ with baseline methods and Section \ref{sec:marked_vs_accum} demonstrates the further advantage of \NonLinearSTHawkes. Section \ref{sec:exp_hawkes} then shows the scalability of our methods on a significantly larger region, where Section \ref{sec:exp_UQ} further examines the empirical coverage of prediction sets by the \rev{CP} method. Finally, Section \ref{sec:conclude} concludes the work with discussion on future steps. The appendix contains additional derivations and algorithms.

\subsection{Related work}

Wildfire prediction and modeling is an essential procedure for analyzing the occurrence of wildfire events. There have many indices, such as the \rev{BI} \cite{Schoenberg2007ACA} and the fire danger index \cite{Sanabria2013SpatialIO} for general awareness of fire risks. Despite their popularity, these indices often fail to account for events' interactions. Meanwhile, regression-based approaches \cite{Frandsen1997IgnitionPO,Plucinski2008LaboratoryDO,Jain2020ARO} are more flexible and often yield satisfactory predictions. However, their performance can be sensitive to the number of available observations per location and thus not applicable under arbitrary spatial granularity with a fixed amount of training data. Lastly, stochastic point-process models \cite{Cunningham1973ASM,Xu2011PointPM,Koh2021SpatiotemporalWM} have been leveraged to examine the conditional fire risk given past data and allow a deeper understanding of the underlying stochastic mechanism. However, most current works focus on model evaluation through the \revtwo{akaike information criterion} (AIC) rather than predicting the binary occurrence of wildfire events using one-class data. In practice, making a binary prediction is essential for forestry managers and utility owners to understand the fire risk.


Since our proposed fire occurrence model is based on the Hawkes process, we briefly survey existing methods in a wider context. Initially proposed in \cite{Hawkes1971SpectraOS}, the Hawkes process is a stochastic temporal point-process model for rates of events conditioning on historical ones. There have been many extensions that take into account spatial interactions \cite{HawkesSpatial1,HawkesSpatial2,HawkesSpatial3} and influences by auxiliary features (i.e., marks) \cite{HawkesMarks1,HawkesMarks2,HawkesMarks3}. Neural-network-based Hawkes process models \cite{HawkesNeural1,HawkesNeural2,HawkesNeural3} have also been proposed for greater model expressiveness. These models have shown great promise in fields such as financial markets \cite{HawkesFinance}, social networks \cite{HawkesSocial}, disease modeling \cite{HawkesDisease}, and neurophysiological studies \cite{HawkesNeural}. Despite their emerging popularity and flexibility, how to make a prediction based on rate estimates and comparisons against predictive models has been less well studied.

We briefly surveyed \rev{CP}, the primary tool used for constructing prediction sets that quantify uncertainty in fire magnitude prediction. Originated in the seminal work \cite{conformaltutorial}, \rev{CP} has gained wide popularity for uncertainty quantification \cite{Zeni2020ConformalPA}. It is particularly appealing as the methods are distribution-free, model-agnostic, and easily implementable. The only assumption is that observations are exchangeable (e.g., i.i.d.). On a high level, \rev{CP} methods assign non-conformity scores to potential outcomes of the response variable. The outcomes that have small non-conformity scores are included in the prediction set. Many methods follow this logic with promising results \cite{Candes_classification,MJ_classification,Eklund2013TheAO, Bosc2019LargeSC, Smith2014AnomalyDO}. More recently, works have also relaxed the exchangeability assumption \cite{CPcovshift,Park2021PACPS,EnbPI,Xu2021ConformalAD,Stankeviit2021ConformalTF,Barber2022ConformalPB}, but time-series \rev{CP} methods are still limited, and their applications to wildfire predictions remain largely unexplored.

\section{Sensing for Wildfire and Real-Data Illustration}\label{sec_data}

The latest technology provides multi-modal data for wildfire risk prediction and monitoring. Below, we briefly describe a few common sensing and data collection techniques \cite{calFire,govDetectingWildfire}.
\begin{itemize}
    \item Air patrols: Patrollers typically consist of a pilot and a trained aerial observer. To identify and report observed wildfire phenomena, the plane flies over predetermined areas during periods associated with elevated fire danger. Wildfire activities are also commonly reported by commercial or recreational pilots. 
    
    \item Infrared technology: Thermal imaging technology is commonly used to detect fire risks hot spots. It is also used to detect wildfire progression, contour the fire impact, and identify residual fire during extinguishment.
    
    \item Computer technology: Various management systems are used to obtain well-rounded multi-modal information. Such systems obtain up-to-date weather information, predict the fire probability and spread rate, and reports moisture levels in the natural surrounding.
\end{itemize}

A feature of our work is that we validate our model on a large-scale multi-modal dataset, 2014--2019 fire incident data collected by the California public utilities commission \cite{calFire}. The wildfire occurrence dataset is publicly available and associated with three large utility companies: PG\&E, SCE, and SDG\&E. A total of 3191 fire incidents are recorded, where the latitude-longitude coordinates of each incident are enclosed within the coordinate rectangle $[32.24,-124,38]\times[41.28,-114.67]$. 

The wildfire data is multi-modal and collecting using various sensing mechanism. Each incident is multi-modal with additional information, which we call \textit{marks} in our model. Marks can be categorized as being discrete/continuous and dynamic/static. Static marks do not change at a given location, and all discrete marks are one-hot encoded to be utilized in the model. Static and discrete marks include (1) existing vegetation type and physiology (EVT\_PHYS)\cite{landfireLANDFIREProgram}, such as the road condition and agricultural condition, (2) the name of the three utility companies, and (3) the fire threat zone, which is classified into three levels indicating increasing levels of static fire danger \cite{calFire}. Dynamic and discrete marks include seasonal information (e.g., spring, summer, autumn, and winter). Dynamic and continuous marks include (1) relative humidity in \% of the surrounding \cite{ucarNLDASNorth} (2) temperature in celsius \cite{ucarNLDASNorth} (3) large fire probability (LFP) \cite{usgsFireDanger}, and (4) fire potential index (FPI) \cite{usgsFireDanger}. In particular, LFP and FPI are forecasted by the \revtwo{United States geological survey} (USGS) to indicate the risks associated with a region. 

To pre-process the multi-modal data, we interpolate missing entries of each continuous mark using the spline function with degree 5. Each feature is also standardized to have unit variance and zero mean and further scaled to lie within the interval $[0,1]$ so that estimated parameters for different marks are on the same scale. The unit for risk prediction is in days, while we allow fractional time values during training where the exact hour and minutes are recorded along each incident.

\vspace{0.1in}

\section{Wildfire Prediction Framework}\label{sec:mtd}

\subsection{Wildfire risk prediction: Mutually exciting spatio-temporal point processes}\label{sec:mtd_hawkes}

We observe a sequence of $n$ fire incidents over a time horizon $[0, T]$, where each observation consists of time $t_i$, location $u_i$, and a mark $m_i \in \mathbb R^p$ (where $p$ is the number of features):
\begin{equation}\label{data}
    x_i=(t_i,u_i,m_i),\ i=1,\ldots,n.
\end{equation}
Note that we specify $u_i \in \{1,\ldots,K\}$ for $K$ locations under space discretization. 

We model these event data using a marked spatio-temporal Hawkes process. Given the $\sigma$-algebra $\mathcal{H}_t$ that denotes all historical fire occurrence before time $t$, the conditional intensity function is the probability of an event occurring at time $t$ and location $k$, with current mark $m$:
\begin{equation}\label{hawkes_def}
    \lambda(t,k,m|\mathcal{H}_t)=\lim _{\Delta t, \Delta u \rightarrow 0} \frac{\mathbb{E}\left[\mathbb N([t, t+\Delta t) \times \rev{B(k, \Delta k)}\times B(m, \Delta m)) \mid \mathcal{H}_{t}\right]}{\Delta t \times \rev{B(k, \Delta k)}\times B(m, \Delta m)},
\end{equation}
where $B(a,r)$ is a ball centered at $a$ with radius $r$ and $\mathbb N$ is the counting measure. 
For notation simplicity, we drop $\mathcal{H}_t$ in (\ref{hawkes_def}) from now on. 

We can use the conditional intensity function above \eqref{hawkes_def} to quantify the fire risk. For mutually exciting point processes, the conditional intensity function depend on the past events and they typically increase the chance of a future event in the neighborhood. This {\it mutual excitation} can be modeled by representing the conditional intensity function \eqref{hawkes_def} as (see, e.g., \cite{reinhart2017review}):
\begin{align}
    \lambda(t,k,m) &= \lambda_g(t,k)f(m|t,k)\nonumber\\ &=\left(\mu(k)+\underset{j:t_j<t}{\sum} \mathcal K(u_j,k,t_j,t)\right) \rev{f(m|t,k)}\label{mod1}, 
\end{align}
which factors the conditional intensity into product of ground process $\lambda_g(t,k)$ and conditional density $f(m|t,k)$. In (\ref{mod1}), $\mu(k)$ is the scalar baseline intensity and $\mathcal K(u_j,k,t_j,t)$ measures spatial and temporal influence from event happening at $t_j$ in $u_j$ till current time $t$ through a kernel function

In general, functions $\mu(k), \mathcal K(u_j,k,t_j,t),$ and $\rev{f(m|t,k)}$ can take many possible forms. 
Such choices often depend on the application of interest. For computation simplicity and model interpretability, here we parametrize the model in (\ref{mod1}) as 
\begin{equation}\label{param_f}
    \mu(k) = \mu_k, \quad \mathcal K(u_j,k,t_j,t)=\alpha_{u_j,k}\beta e^{-\beta(t-t_j)}. 
\end{equation}
\rev{In equation \eqref{param_f}, the parameters $\mu_k$ represent the baseline rate of fire risk at location $k$. The parameters $\alpha_{u_j,k}$ capture the spatial influence of fire incidents that occurred at location $u_j$ and time $t_j$ on the fire risk at location $k$ and time $t$. To simplify the design of $\mathcal K(u_j,k,t_j,t)$ in \eqref{param_f}, we use a negative exponential model. This choice is motivated by two key factors. Firstly, it results in an optimization problem whose parameters can be efficiently estimated with a performance guarantee (refer to Section \ref{theory}). Secondly, domain experts have observed that past fire incidents can affect the risk of future fire incidents, but the impact of past events diminishes quickly over time.}

Furthermore, we assume the distribution of the mark is either in linear form or, more generally, through a non-linear function $g$ 
\begin{align}
    & \rev{f(m|t,k)}=\gamma^T m, &&  (\mathtt{LinearSTHawkes}) \label{eq:param_marks} \\
    & \rev{f(m|t,k)}=\rev{g(m|t,k)} &&  (\mathtt{NonLinearSTHawkes}) \label{eq:feature_extractor}
\end{align}
\rev{Even though \eqref{eq:param_marks} is linear, it implicitly incorporates the spatial-temporal information through the mark $m$, which is collected in location $k$ at time $t$.} Meanwhile, \rev{$g(m|t,k)$ in \eqref{eq:feature_extractor}} can \rev{be} any feature extractor (e.g., neural networks) that outputs the score of $m$. \rev{Regarding the formulation differences of \eqref{eq:param_marks} and \eqref{eq:feature_extractor}, }note that \LinearSTHawkes \ based on \eqref{eq:param_marks} is more interpretable, and also leads to more computationally efficient sequential convex optimization scheme with guarantees (see Section \ref{theory_param_recovery}). On the other hand, \NonLinearSTHawkes \ can be more expressive \rev{in terms of capturing the dependency of fire risks on marks through the feature extractor $g(m|t,k)$ in \eqref{eq:feature_extractor}.}

\subsection{Point process parameter estimation and real-time prediction}\label{sec:mtd_hawkes_est}

We estimate the parameters in the model through maximum likelihood. For \LinearSTHawkes, denote all parameters using $\theta=\{\mu,A,\beta,\gamma\}$, where $\mu=\{\mu_k\}_{k=1}^K$ and $A=[\alpha_{i,j}]_{i,j=1}^{K}$. 
We can derive and simplify the log-likelihood of $x_1,\ldots,x_n$ as follows similar to \cite{reinhart2017review} \rev{(the full derivation can be found in appendix \ref{log-lik-deriv})}: 
\begin{align}
    \ell(\theta) =
    & \sum_{i=1}^n \log(\lambda_g(t_i,u_i))+ \sum_{i=1}^n \log(\rev{f(m_i|t_i,u_i)})-\sum_{k=1}^K\int_0^T \lambda_g(\tau,k)d\tau \nonumber\\= 
    & \sum_{i=1}^n \log\left(\mu(u_i) +\underset{j:t_j<t_i}{\sum}\alpha_{u_j,u_i}\beta e^{-\beta(t_i-t_j)}\right)\nonumber + \sum_{i=1}^n\log(\rev{f(m_i|t_i,u_i)})\\
     &~~~ - \sum_{k=1}^K T\mu(k) - \sum_{i=1}^n\left(\sum_{k=1}^K  \alpha_{u_i,k}\right)(1-e^{-\beta(T-t_i)}).\label{loglik}
\end{align}
Note that the likelihood term of the marks decouples from the rest. Thus, when using \NonLinearSTHawkes \ based on \eqref{eq:feature_extractor}, we first fit a feature extractor on the  marks and then employ maximum likelihood estimation to estimate the rest parameters. 
To achieve better model estimation stability (since we believe few features should be effective in the model), we further add $\ell_1$ regularization on $\gamma$: 
\begingroup
\allowdisplaybreaks
\begin{align}
\min_{\theta=\{\mu,A,\beta,\gamma\}}
\quad - & \sum_{i=1}^n \log\left(\mu(u_i) +\underset{j:t_j<t_i}{\sum}\alpha_{u_j,u_i}\beta e^{-\beta(t_i-t_j)}\right)\nonumber - \sum_{i=1}^n\log(\gamma^Tm_i)\\
    + & \sum_{k=1}^K T\mu(k) + \sum_{i=1}^n\left(\sum_{k=1}^K  \alpha_{u_i,k}\right)(1-e^{-\beta(T-t_i)}) + \| \gamma\|_1  \label{mod} \\
\textrm{subject to} 
\quad & \alpha_{i,j}=0 \textrm{ if } |i-j|\geq\tau, \label{sparse_influ} \\
\quad & \|\mu\|_2\leq 1, \|A\|_2\leq 1, \|\gamma\|_2\leq 1,\label{bounded_para}\\
\quad & \beta\geq 0, \mu(u_i) \geq 0 \ \forall u_i. \label{nonneg}
\end{align}
\endgroup
The purpose of constraints \eqref{sparse_influ}-\eqref{nonneg} can be explained as follows: (\ref{sparse_influ}) introduces sparsity in the interaction matrix and reduces the total number of parameters in the model for computational efficiency; (\ref{bounded_para}) ensures the objective (\ref{mod}) is bounded and is reasonable since the rate $\lambda(t,k,m)$ is typically very small; (\ref{nonneg}) is introduced since baseline rates (i.e. $\mu(u_i)$) and interaction propagation over time (i.e. $\beta$) are non-negative. Note that the constraints define a convex feasible region.

In addition, we can show that $\ell(\theta)$ is concave in all other parameters with a fixed scalar $\beta$. Thus, we can device a method to solve (\ref{mod}) to global optimal solution: for a grid of $\beta$ values, solve the corresponding convex optimization problem using solvers such as \cite{diamond2016cvxpy} to high numerical accuracy, and then choose the optimal $\beta$ that gives the best overall objective value. The description of the algorithm\rev{, as well its computational efficiency, } is in Algorithm \ref{alter_algo} \rev{of Appendix \ref{appendix:altermin}}. 
In our experiments, we observe that the algorithm usually terminates in a small number of iterations (e.g., three), and each iteration only takes a few seconds to minutes, depending on the problem size. Hence, it is computationally friendly. 

\subsection{Fire magnitude prediction: Conformal prediction set}\label{sec:mtd_conformal} \label{sec:conformal}

Besides predicting when and where fire occurs, fire magnitude prediction is also desirable ---knowing the possible fire magnitude can better inform decision-makers of potential losses by such disasters and plan accordingly. The dataset described in Section \ref{sec_data} treats fire magnitude as discrete categories in its catalog. In principle, this can thus be achieved by variants of \LinearSTHawkes \ and \NonLinearSTHawkes \ for categorical data. However, making categorical prediction based on the estimated risks requires us to construct multi-class thresholds, which can greatly increase model design complexity. In addition, it is unclear how to quantify uncertainty in the resulting categorical estimates.

Thus, we treat fire magnitude prediction as a classification problem: given multi-modal features $X_i\in \R^p$ as in \eqref{data}, we would like to build a multi-class classifier that outputs $\hat{Y}_i \in \{1,\ldots,C\}$ as the fire magnitude prediction (assuming $C$ magnitude levels). Denote $\pi_i:=P_{Y_i|X_i}$ as the true conditional distribution of $Y_i|X_i$, whose properties are unknown. In a typical classification setting, we assume the first $N$ data are known to us as training data and the goal is to construct an estimator $\hat{\pi}:=\A(\{(X_i,Y_i)\}_{i=1}^N)$, which satisfies $\sum_{c=1}^C \hat{\pi}_{X_i}(c)=1, \hat{\pi}_{X_i}(c)\geq 0$ for any $i\geq 1$. Here, $\A$ is any classification algorithm, from the simplest multinomial logistic regression to a complex deep neural networks. Then, the point prediction $\hat{Y}_i:=\arg \max_{c\in [C]} \hat{\pi}_{X_i}(c)$ is obtained for any test index $i>N$.

However, point predictions are often insufficient in such settings---there are inherent uncertainties in these predictions, which arise due to randomness in data generation, during the collection of multi-modal data, and when fitting the multi-class classifier. Therefore, a \textit{confident} fire magnitude prediction is essential, which quantifies uncertainties in the point predictions and contains all the possible high-probability outcomes. One way for uncertainty quantification in classification is the construction of \textit{prediction sets} around $\hat{Y}_i$ that contain actual observations $Y_i$ with high probability before its realization. Formally, given a significance level $\alpha$ \rev{$\in (0,1)$}, we construct a \textit{prediction set} $\widehat{C}(X_i, \alpha) \subset \{1,\ldots,C\}$ such that
\begin{equation}\label{coverage}
    \mathbb{P}(Y_i\in \widehat{C}(X_i, \alpha)) \geq 1-\alpha.
\end{equation}
\rev{We note that the significance level $\alpha$ in conformal prediction should be distinguished from the interaction parameters $\alpha_{ij}$ in the point-process model, the latter of which has double subscripts as in \eqref{param_f}.}
A set satisfying \eqref{coverage} thus confidently predicts the actual fire magnitude $Y_i$ with high probability. Note that a trivial construction that always satisfies \eqref{coverage} is $\widehat{C}(X_i, \alpha) = \{1,\ldots,C\}$, so we also want the prediction set to be as small as possible. This is a challenging question because fire incidents are highly correlated and non-stationary, and classifiers can be very complex (e.g., neural network classifiers).

To build prediction sets that satisfy \eqref{coverage} in practice, we produce uncertainty sets using recent advances in \rev{CP} \cite{MJ_classification,EnbPI,xuERPAS2022}. CP methods requires two ingredients. \rev{First, they define} \textit{non-conformity scores}, which quantify the dissimilarity of a potential fire magnitude. \rev{Second, they specify} the prediction set based on non-conformity scores. \rev{As a result,} CP methods assign non-conformity scores to each possible fire magnitude and the prediction set \rev{contains} fire magnitude whose non-conformity scores are small compared to past ones.

We first specify a particular form of non-conformity score recently developed in \cite{MJ_classification} using any estimator $\hat{\pi}$. The notations are very similar and we include the descriptions for a self-contained exposition. Given the estimator $\hat{\pi}$, for each possible label $c$ at test feature $X_i, i>N$, we make two other definitions:
\begin{align}
 & m_{X_i}(c):=\sum_{c'=1}^C \hat{\pi}_{X_i}(c')\cdot \mathbb{I}(\hat{\pi}_{X_i}(c')>\hat{\pi}_{X_i}(c)).\label{def:tot_mass}\\
   & r_{X_i}(c):=\left \vert\sum_{c'=1}^C \mathbb{I}(\hat{\pi}_{X_i}(c')>\hat{\pi}_{X_i}(c))\right\vert+1. \label{def:rank}
\end{align}
where $\mathbb I$ is the indicator function. 
In other words, (\ref{def:tot_mass}) calculates the total probability mass of labels deemed more likely than $c$ by $\hat{\pi}$. It strictly increases as $c$ becomes less probable. Meanwhile, (\ref{def:rank}) calculates the rank of $c$ within the order statistics. It is also larger for less probable $c$. Given a random variable $U_i\sim \text{Unif}[0,1]$ and pre-specified regularization parameters $\{\lambda, \kreg\}$, we define the non-conformity score as
\begin{equation}\label{eq:tau_MJ}
    \hat{\tau}_i(c):=m_{X_i}(c)+\underbrace{\hat{\pi}_{X_i}(c)\cdot U_i}_{(i)} + \underbrace{\lambda(r_{X_i}(c)-\kreg)^+}_{(ii)}.
\end{equation}
We interpret terms (i) and (ii) in (\ref{eq:tau_MJ}) as follows. Term (i) randomizes the uncertainty set, accounts for discrete probability jumps when new labels are considered. A similar randomization factor is used in \cite[Eq. (5)]{Candes_classification}. In term (ii), $(z)^+:=\max(z,0)$. Meanwhile, the regularization parameters $\{\lambda, \kreg\}$ force the non-conformity score to increase when $\lambda$ increases and/or $\kreg$ decreases. In words, $\lambda$ denotes the additional penalty when the label is less probable by one rank and $\kreg$ denotes when this penalty takes place. This term ensures that the sets are \textit{adaptive}, by returning smaller sets for easier cases and larger ones for harder cases. 

Then, the prediction set based on \eqref{eq:tau_MJ} is
\begin{equation}\label{eq:set}
    \widehat{C}(X_i,\alpha):=\{c\in [C]:  \sum_{j={i-N}}^{i-1} \mathbb{I}(\hat \tau_j\leq \hat{\tau}_i(c))/N < 1-\alpha\},
\end{equation}
where $\hat \tau_j:= \hat{\tau}_j(Y_j)$. The set in \eqref{eq:set} includes all the labels whose non-conformity scores are no greater than $(1-\alpha)$ fraction of previous $N$ non-conformity scores. Following \eqref{eq:tau_MJ} and \eqref{eq:set}, we thus propose \revtwo{\textit{ensemble regularized adaptive prediction set}} (\ERAPS) in Algorithm \ref{algp:eraps}. In particular, \ERAPS \ aggregates probability predictions from bootstrap multi-class classifiers to yield more accurate point prediction and leverage new feedback of $Y_i$ to ensure adaptiveness in the prediction sets.
\begin{algorithm}[t]
\cprotect \caption{Ensemble Regularized Adaptive Prediction Set}
\label{algp:eraps}
\begin{algorithmic}[1]
\Require{Training data$\{(X_i, Y_i)\}_{i=1}^N$, classification algorithm $\mathcal{A}$, $\alpha$, regularization parameters $\{\lambda,\kreg\}$, aggregation function $\phi$ \rev{(e.g., mean)}, number of bootstrap models $B$, the batch size $s$, and test data $\{(X_i,Y_i)\}_{i=N+1}^{N+N_1}$, with $Y_i$ revealed only after the batch of $s$ prediction intervals with $i$ in the batch are constructed.}
\Ensure{Ensemble uncertainty sets $\{\widehat{C}(X_i,\alpha)\}_{i=N+1}^{N+N_1}$}
\For {$b = 1, \dots, B$}  \Comment{Train Bootstrap Estimators}
\State Sample with replacement an index set $S_b=(b_1,\ldots,b_N)$ from indices $(1,\ldots,N)$.
\State Compute $\hat{\pi}^b=\mathcal{A}(\{(X_i,Y_i) \mid i \in S_b \})$.
\EndFor
\State Initialize $\boldsymbol \tau=\{\}$ and sample $\{U_i\}_{i=1}^{N+N_1}\overset{i.i.d.}{\sim}\text{Unif}[0,1]$.
\For {$i=1,\dots,N$} \Comment{LOO Ensemble Estimators and Scores}
\State Compute $\hat{\pi}_{-i}^{\phi}:=\phi(\{\hat{\pi}^b: i\notin S_b \})$ such that for each $c\in \{1,\ldots,C\}$
\Statex \hskip \algorithmicindent$\hat{\pi}_{-i,X_i}^{\phi}(c)=\phi(\{\hat{\pi}^b_{X_i}(c): i \notin S_b \}).$
\State Compute $\hat \tau_i^{\phi}:=\hat \tau_{X_i}(Y_i)$ using (\ref{eq:tau_MJ}) and $\hat{\pi}_{-i}^{\phi}$. 
\State $\boldsymbol \tau=\boldsymbol \tau \cup \{\hat{\tau}_i^{\phi}\}$
\EndFor
\For {$i=N+1,\dots,N+N_1$} \Comment{Build Uncertainty Sets}
\State Compute $\hat \tau_{i,cal}^{\phi}:=q_{\boldsymbol \tau,1-\alpha}(\boldsymbol \tau)$ as the $(1-\alpha)$-empirical quantile of $\boldsymbol{\tau}$.
\State Compute $\hat \pi_{-i}^{\phi}:=\phi(\{\hat \pi_{-i}^{\phi}\}_{i=1}^N)$ so that for each $c\in \{1,\ldots,C\}$
\Statex \hskip \algorithmicindent $\hat \pi_{-i,X_i}^{\phi}(c):=\phi(\{\hat \pi_{-i,X_i}^{\phi}(c)\}_{i=1}^N).$
\State Compute $\widehat{C}(X_i,\alpha)$ in (\ref{eq:set}) using $\hat \pi_{-i}^{\phi}$ and $\hat \tau_{i,cal}^{\phi}$.
\If {$t-T$ = 0 mod $s$} \Comment{Slide Scores Forward}
\For {$j=i-s,\ldots,i-1$}
\State Compute $\hat \tau_j^{\phi}:=\hat \tau_{X_j}(Y_j)$ using (\ref{eq:tau_MJ}) and $\hat{\pi}_{-j}^{\phi}$.
\State $\boldsymbol\tau=(\boldsymbol\tau- \{\hat{\tau}_1^{\phi}\}) \cup \{\hat{\tau}_j^{\phi}\}$ and reset index of $\boldsymbol\tau$.
\EndFor
\EndIf
\EndFor
\end{algorithmic}
\end{algorithm}

\section{Theoretical Guarantee}\label{theory}

In this section, we establish some theoretical performance guarantees for the proposed algorithms. 
Section \eqref{theory_param_recovery} provides parameter recovery guarantee for the point-process model defined in \eqref{mod1}. Section \eqref{theory_conformal} provides coverage guarantee (see Eq. \eqref{coverage}) and the tightness of the fire magnitude prediction set by \ERAPS.

\subsection{Parameter recovery for point process model}\label{theory_param_recovery}

Note that for fixed $\beta$, the problem for estimating the rest of the parameters in $\theta$ via \eqref{loglik} for \LinearSTHawkes \ is convex (it can be shown that the objective function is concave in $\theta$ other than $\beta$, and constraints induce convex feasible domain). We can establish the following bound using a similar technique as in \cite{juditsky2019signal,zhang2021solar}. We do not consider the bound for \NonLinearSTHawkes \ in \eqref{eq:feature_extractor} because it is impossible to verify convexity for a generic feature extractor $g$.

We first obtain parameter recovery bound for minimizing a generic continuously differentiable strictly convex function $f(\theta):\Theta \rightarrow \R,$ where $\Theta \subset \R^p$ is a convex set. Let $F(\theta):=\nabla f(\theta)$ be the gradient of $f$ on $\Theta$. We know that $F(\theta)$ is \textit{monotone} \cite{juditsky2019signal}:
\[
[F(\theta)-F(\theta')]^T[\theta-\theta'] \geq 0 \ \forall \theta,\theta' \in \Theta.
\]
Let $\theta^*\in \Theta$ be the unique global minimizer of $f$, which exists as $f$ is strictly convex. To estimate $\theta^*$, we use the projected gradient descent procedure, starting at an arbitrary $\theta_0 \in \Theta:$
\begin{equation}\label{pgd}
\theta_k := \text{Proj}_{\Theta}(\theta_{k-1}-t_k F(\theta_{k-1})),
\end{equation}
where $t_k > 0$ determines the step size and $\text{Proj}_{\Theta}(\hat{\theta}):=\arg\min_{\theta \in \Theta} \|\hat{\theta}-\theta\|_2$. To analyze the error $\|\theta_k-\theta^*\|_2$ after $k$ iterations, we need the following conditions:
\textit{\begin{assumption}\label{param_assume}
    Assume that there exist $D,\kappa,M >0$ where
    \begin{align}
         (i) \ &\|\theta-\theta'\|_2 \leq D \ \forall \theta, \theta' \in \Theta, \label{bonded_domain} \\
         (ii) \ &[F(\theta)-F(\theta')]^T[\theta-\theta'] \geq \kappa \|\theta-\theta'\|^2_2 \ \forall \theta, \theta' \in \Theta, \label{strongly_monotone} \\
         (iii) \ &\|F(\theta)\|_2\leq M \ \forall \theta \in \Theta. \label{bounded_grad}
    \end{align}
\end{assumption}}
\rev{We now have the following lemma that yields the error bound in \eqref{err_bound}. The proof is contained in appendix \ref{proof_lem1}.}
\textit{\begin{lemma}\label{para_recovery_lem}
    Under Assumptions \ref{param_assume}:\eqref{bonded_domain}---\eqref{bounded_grad} and with the step sizes 
    \begin{equation}\label{step_size}
        t_k := [\kappa(k+1)]^{-1},
    \end{equation}
    Estimates $\theta_k$ obtained through \eqref{pgd} obey the error bound
    \begin{equation}\label{err_bound}
        \|\theta_k-\theta^*\|_2^2 \leq \frac{M^2}{\kappa^2(k+1)}.
    \end{equation}
\end{lemma}}

We can now use Lemma \ref{para_recovery_lem} to obtain the parameter recovery guarantee for minimizing $\ell(\theta)$ via solving \eqref{loglik}. For a fixed $\beta > 0$, let 
\begin{equation}\label{all_but_beta}
    \thetabeta:=\theta-\{\beta\}
\end{equation}
contain all the model parameters except $\beta$ when solving \eqref{loglik}. We thus know that under Lemma \ref{para_recovery_lem}, the estimate $\hat{\theta}[\beta]$ converges to the global minimum $\theta^*[\beta]$ at rate $1/k$. Meanwhile, since the optimal parameter $\beta^*$ is non-negative scalar, we can estimate it up to arbitrary precision using one one-dimensional grid search. In particular, assume $\beta^*\in [\beta_0,\beta_1]$ with known values of $\beta_0,\beta_1$. For a fixed integer $J\geq 1$, divide the region $[\beta_0,\beta_1]$ into $J+1$ points $\beta_0,\ldots,\beta_{J}$, where
\begin{equation}\label{beta_j}
\beta_j:=\beta_0+\frac{j}{J}(\beta_1-\beta_0), j=0,\ldots,J.
\end{equation}
Then, we can obtain estimates $\hat{\theta}[\beta_j]$ via solving \eqref{loglik} using the projected gradient descent procedure \eqref{pgd} at the fixed $\beta_j$. Given $J$ pairs of estimates $(\beta_{j},\hat{\theta}[\beta_{j}])$, we define
\begin{align}
        \hat{\theta}&:=(\beta_{j^*},\hat{\theta}[\beta_{j^*}]) \label{estimate_smallest} \\ 
        j^*&:=\underset{j=0,\ldots,J}{\arg\min} \ \ell([\beta_j,\hat{\theta}[\beta_j]]), \label{c_star}
\end{align}
which denotes the estimate that reaches the smallest log-likelihood out of these $M$ estimates. \rev{We then bound in the following theorem the parameter estimation error of $\hat{\theta}$ in \eqref{estimate_smallest}. The proof is contained in appendix \ref{proof_thm1}.}
\textit{\begin{theorem}[\LinearSTHawkes \ parameter recovery guarantee]\label{para_recovery_final}
    Let $\theta^*$ be a minimizer of $\ell(\theta)$ in \eqref{loglik} under \LinearSTHawkes \ in \eqref{eq:param_marks}. Under Assumption \ref{param_assume}:\eqref{bonded_domain}---\ref{bounded_grad}, the estimate $\hat{\theta}$ in \eqref{estimate_smallest} obeys the bound
    \begin{equation}\label{bound_final}
        \|\hat{\theta}-\theta^*\|_2^2 = \mathcal{O}\left(\frac{1}{J^2}+\frac{1}{k+1}\right).
    \end{equation}
    In \eqref{bound_final}, $J$ is the number of grid searches for $\beta^*$ in $[\beta_0,\beta_1]$ and $k$ is the number of projected gradient descent step \eqref{pgd} of $\theta[\beta_j]$ in \eqref{all_but_beta} at each search point $\beta_j$.
\end{theorem}}

\vspace{.1in}
\noindent
The implication of Theorem \ref{para_recovery_final} is that we can recover the \textit{true model} of $\lambda(t,k,m)$ in \eqref{mod1} for \LinearSTHawkes \ in \eqref{eq:param_marks}. This is because \LinearSTHawkes \ reaches the smallest negative log-likelihood under $\theta^*$ and log likelihood is also the highest under the true model. Thus, when estimates $\hat{\theta}$ approach true parameters $\theta^*$ in $\ell_2$ norm, the corresponding model estimate also recover the true model.

\subsection{Conformal prediction set guarantee}\label{theory_conformal}

Note that in existing \rev{CP} literature, it is typically assumed that observations $(X_i,Y_i)$ are exchangeable. This assumption is unrealistic in our setting when strong correlation exists within data. Instead, we impose assumptions on the quality of estimating the non-conformity scores and on the dependency of non-conformity scores in order to bound coverage gap of (\ref{coverage}). Most of the assumptions and proof techniques extends our earlier work \cite{EnbPI}, but we extend it to the classification setting under arbitrary definitions of non-conformity scores. In particular, we allow arbitrary dependency to exist within features $X_i$ or responses $Y_i$.

Given any feature $X$, a possible label $c$, and a probability mapping $p$ such that $\sum_{c=1}^C p_X(c)=1, p_X(c)\geq 0$, we denote $G:(X,c,p)\rightarrow \mathbb{R}$ as an arbitrary non-conformity mapping and $\tau^p_X(c):=G(X,c,p)$ as the non-conformity score at label $c$. For instance, we may consider 
\begin{equation}\label{ex:NCM}
    G(X,c,p)=\sum_{c'=1}^C p_{X}(c')\cdot \mathbb{I}\{p_{X}(c')>p_{X_i}(c)\},
\end{equation}
which computes the total probability mass of labels that are deemed more likely than $c$ by $p$. The less likely $c$ is, the greater $\tau^p_i(c)$ is, indicating the non-conformity of label $c$. For notation simplicity, the oracle (resp. estimated) non-conformity score of each training datum $(X_i,Y_i), i=1,\ldots,N$ under the true conditional distribution $\pi:=P_{Y|X}$ (resp. any estimator $\hat{\pi}$) is abbreviated as $\tau_i=\tau^{\pi}_{X_i}(Y_i)$ (resp. $\hat \tau_i$). 

We now impose these two assumptions that are sufficient for bounding coverage gap of (\ref{coverage}). \rev{First, we make assumptions about the quality of estimation by the chosen classifier:}
\textit{\begin{assumption}[Error bound on estimation]\label{assumption:estimation}
Assume there is a real sequence $\{\vartheta_i\}$ where
$\frac 1 N \sum_{j={i-N}}^{i-1} (\hat \tau_j-\tau_j)^2\leq \vartheta_N^2.$
\end{assumption}}
\rev{Then we make assumptions about to the property of true non-conformity scores:}
\textit{\begin{assumption}[Regularity of non-conformity scores]\label{assumption:regularity}
Assume $\{\tau_j\}_{j={i-N}}^i$ are independent and identically distributed (i.i.d.) according to a common cumulative density function (CDF) $F$ with Lipschitz continuity constant $L>0$.
\end{assumption}}

We brief remark on implications of the Assumptions above. Note that Assumption \ref{assumption:estimation} essentially reduces to the point-wise estimation quality of $\pi$ by $\hat{\pi}$, which may fail under data overfitting---all $N$ training data are used to train the estimator. In this case, $\hat{\pi}$ tends to over-concentrate on the empirical conditional distribution under $(X_i,Y_i), i=1,\ldots,N$, which may not be representative of the true conditional distribution $P_{Y|X}$. A common way to avoid this in the CP literature is through data-splitting---train the estimator on a subset of training data and compute the estimated non-conformity scores $\hat{\tau}$ only on the rest training data (i.e., calibration data). However, doing so likely results in a poor estimate of $\pi$ and as we will see, the theoretical guarantee heavily depends on the size of estimated non-conformity scores. On the other hand, Assumption \ref{assumption:regularity} can be relaxed as stated in \cite{EnbPI}. For instance, the oracle non-conformity scores can either follow linear processes with additional regularity conditions \cite[Corollary 1]{EnbPI} or be strongly mixing with bounded sum of mixing coefficients \cite[Corollary 2]{EnbPI}. The proof techniques directly carry over, except for slower convergence rates. 

Lastly, define the empirical distributions using oracle and estimated non-conformity scores:
\begin{align*}
    & \tildeF{x}:=\frac 1 N\sum_{j={i-N}}^{i-1} \mathbb{I}(\tau_j\leq x), && \text{[Oracle]}\\
    & \hatF{x}:= \frac 1 N\sum_{j={i-N}}^{i-1} \mathbb{I}(\hat \tau_j\leq x). && \text{[Estimated]}
\end{align*}

We then have the following coverage results at the prediction index $t>T$. 
\textit{\begin{lemma}[\rev{\cite[Lemma 2]{EnbPI}}]\label{lem:tildeFandhatF}
Suppose Assumptions \ref{assumption:estimation} and \ref{assumption:regularity} hold. Then,
\[
\sup_{x}|\tildeF{x}-\hatF{x}|\leq (L+1)\vartheta_N^{2/3}+2 \sup_{x}|\tildeF{x}-F(x)|.
\]
\end{lemma}}
\rev{The proof of Lemma \ref{lem:tildeFandhatF} appears in Appendix \ref{proof_lem2}.}
\textit{\begin{lemma}[\rev{\cite[Lemma 1]{EnbPI}}]\label{lem:tildeFandF}
Suppose Assumption \ref{assumption:regularity} holds. Then, for any training size $N$, there is an event $A$ within the probability space of non-conformity scores $\{\tau_j\}_{j=1}^N$, such that when $A$ occurs,
\[
    \sup_{x}|\tildeF{x}-F(x)|\leq \sqrt{\log(16N)/N}.
\]
In addition, the complement of event $A$ occurs with probability $\mathbb P(A^C)\leq \sqrt{\log(16N)/N}.$
\end{lemma}}
\rev{The proof of Lemma \ref{lem:tildeFandF} appears in Appendix \ref{proof_lem3}.}

As a consequence of Lemmas \ref{lem:tildeFandhatF} and \ref{lem:tildeFandF}, the following bound of coverage gap of (\ref{coverage}) holds:
\textit{\begin{theorem}[Coverage guarantee, \rev{\cite[Theorem 1]{EnbPI}}]\label{thm:asym_cond_cov}
Suppose Assumptions \ref{assumption:estimation} and \ref{assumption:regularity} hold. For any training size $N$ and significance level $\alpha\in (0,1)$, we have 
\begin{equation}\label{eq:guarantee}
    |\PP(Y_i \notin \widehat{C}(X_i,\alpha))-\alpha|\leq 24\sqrt{\log(16N)/N}+4(L+1)\vartheta_N^{2/3}.
\end{equation}
\end{theorem}}
\rev{The proof of Theorem \ref{thm:asym_cond_cov} appears in Appendix \ref{proof_thm2}.} Note that Theorem \ref{thm:asym_cond_cov} holds uniformly over all $\alpha \in [0,1]$ because Lemmas \ref{lem:tildeFandhatF} and \ref{lem:tildeFandF} bound the sup-norm of differences of distributions. Hence, users in practice can select desired parameters $\alpha$ \textit{after} constructing the non-conformity scores. Such a bound is also useful when building multiple prediction intervals simultaneously, under which $\alpha$ is corrected to reach nearly valid coverage \cite{Farcomeni2008ARO}.

In addition to coverage guarantee, we can analyze the convergence of $\widehat{C}(X_i,\alpha)$ to the oracle prediction set $C^*(X_i,\alpha)$ under further assumptions. Given the true conditional distribution function $\pi:=P_{Y|X}$, we first order the labels so that $\pi_{X_i}(i)\geq \pi_{X_i}(j)$ if $i\leq j$. Then, we have
\[
C^*(X_i,\alpha)=\{1,\ldots,c^*\},
\]
where $c^*:=\min_{c\in [C]} \sum_{k=1}^c \pi_{X_i}(k)\geq 1-\alpha.$
\begin{theorem}[Set size convergence guarantee]\label{thm:asy_set}
\textit{Suppose Lemmas \ref{lem:tildeFandhatF} and \ref{lem:tildeFandF} hold and denote $F^{-1}$ as the inverse CDF of $\{\tau_j\}_{j=i-N}^i$. Further assume that }
\begin{itemize}
\item[(1)] \textit{$c^*_1=c^*_2$ where}
\begin{align*}
    c^*_1&:= \arg\min_{c} \left\{\sum_{k=1}^c \pi_{X_i}(k) \geq 1-\alpha\right\}, \\
    c^*_2&:=\arg\max_{c} \left\{\tau_i(c) < F^{-1}(1-\alpha)\right\}.
\end{align*}
\item[(2)] \textit{There exists a sequence $\vartheta'_i$ converging to zero with respect to $N$ such that $\|\tau_i-\hat{\tau}_i\|_{\infty} \leq \vartheta'_i$, where the $\infty$-norm is taken over class labels.}
\end{itemize}
\textit{Then, there exists $N$ large enough such that for all $i > N$,}
\begin{equation}\label{set_diff}
    \widehat{C}(X_i,\alpha) \Delta C^*(X_i,\alpha) \leq 1,
\end{equation}
\textit{where $\Delta$ in \eqref{set_diff} denotes set difference.}
\end{theorem}
\rev{The proof of Theorem \ref{thm:asy_set} appears in Appendix \ref{proof_thm3}.} Note that if the non-conformity score at any label $c$ is defined in \eqref{ex:NCM}, which is the total probability mass of labels $c'\neq c$ that are more likely than $c$ based on a conditional probability mapping $p$, then the first additional assumption (i,e., $c^*_1=c^*_2$) in Theorem \ref{thm:asy_set} can be verified to hold. In general, whether this assumption is satisfied depends on the particular form of the non-conformity score.

\section{Model Validation by Real-Data}\label{sec:exp}

We apply the proposed models on the 2014-2019 California wildfire data described in Section \ref{sec_data}. The experiment is organized as follows. Section \ref{sec:setup} describes the setup details, including the dataset and evaluation metrics. Section \ref{sec:exp_Hawkes_carved} compares \LinearSTHawkes \ with competing baselines on data from a small region. Section \ref{sec:marked_vs_accum} compares \LinearSTHawkes \ and \NonLinearSTHawkes \ on the same region to highlight their performance differences.

\subsection{Evaluation metrics} \label{sec:setup} 

We use the $F_1$ score for performance assessment, which is a standard metric for classification when data are \textit{imbalanced}---note that the number of no occurrence of fire incidents (denoted as 0) significantly outweighs the other (denoted as 1). The goal is to predict as many fire occurrences as possible without making too many false positives. \rev{In our case, false positives measured at each location refers to be a prediction of fire incidents at a specific date $t$ when there is no fire incident.} Quantitatively, we define the set of fire occurrences as $U$ and our predicted set as $V$. Then the \textit{precision} $P$ and \textit{recall} $R$ are defined as
\begin{equation}\label{P_and_R}
    P=|U\cap V|/|V|, \ R=|U\cap V| / |U|,
\end{equation}
where the notation $|\cdot|$ denotes the size of the set. \rev{In the definition \eqref{P_and_R}, we write $P$ and/or $R$ to be 1 if the ratio is 0/0 (i.e., there is no fire incident at a specific location and the model correct predicts none).} The $F_1$ score is thus a combination: $F_1\rev{=2/(P^{-1}+R^{-1})}=2PR/(P+R)$, where a high $F_1$ score indicates both a large of true detection and a small number of false positives. \rev{In general, when one of $P$ and $R$ is more important, one can consider a weighted $F_1$ that assigns imbalanced weights to precision and recall. We use non-weighted $F_1$ scores in all our experiments.}

We construct dynamic thresholds to make binary prediction based on estimated fire risk $\hat{\lambda}(t,k,m)$ defined in Eq. \eqref{mod1}. The detailed Algorithm \ref{algo:thres} is provided in appendix \ref{appendix:thres}.
In particular, we observe that rate estimates $\hat{\lambda}(t,k,m)$ have clear seasonality (e.g., a sharp drop from summer to fall and a sharp rise from spring to summer). At the same time, fire incidents often occur when rate estimates suddenly increase on certain days. For instance, Figure \ref{fig:loc_pred_carved} illustrates the performance of our model based on the observations above. 

\begin{figure}[!t]
    \includegraphics[width=\linewidth]{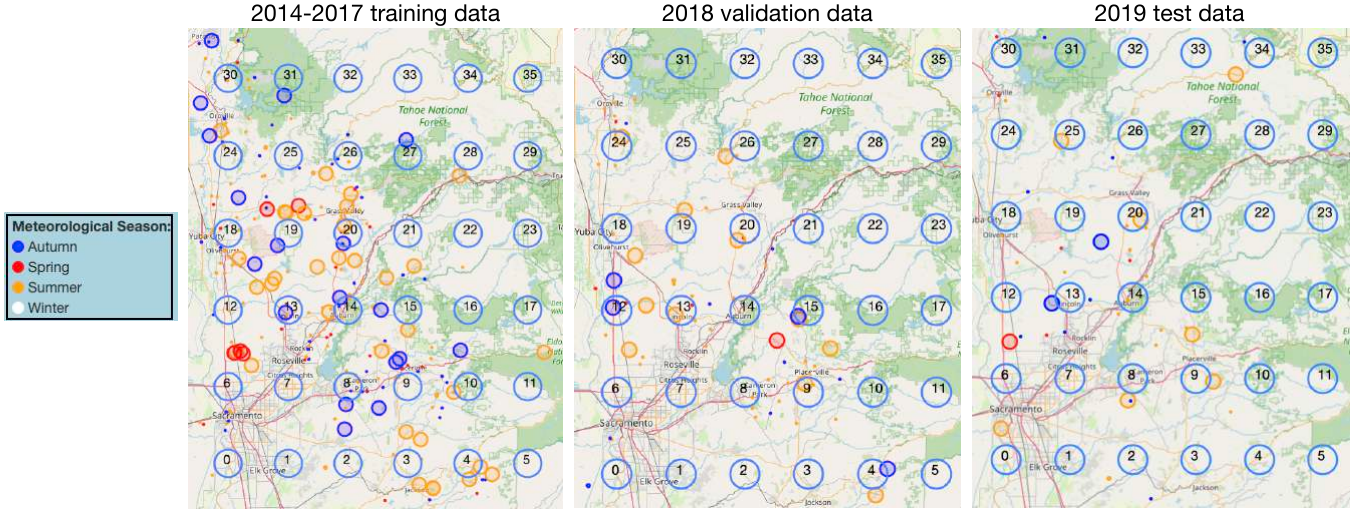}
\caption{Visualize data and grid discretization on data from different years. There are grid-wise shifts in data distribution---for instance, fire incidents cluster more closely around grid 12 in 2018 (validation) than in 2014---2017 (training) or in 2019 (test).}
\label{true_data_small}
\end{figure}

\subsection{LinearSTHawkes vs. Baselines} \label{sec:exp_Hawkes_carved}

We first focus on a small region because the distribution of fire incidents within the region and the performance of our model can be visualized clearly. The model is trained with incidents between 2014 and 2017 and examined on validation data in 2018. There were 238 fire occurrences in 2014-2017 and 70 in 2018. Upon consulting domain experts, we set the sides of discretized cells to be 0.24-degree in both longitude and latitude directions so that 36 non-overlapping cells cover the region. Figure \ref{true_data_small} visualizes both the training and validation data, from which it is clear that the validation data have a much less number of actual fires; only a few grids have fires that occurred near them.

\vspace{0.1in}
\noindent \textit{Estimated parameters.} In practice, our feature $m_i$ includes both temporal dynamic features $m_d$ (e.g., weather information) and location-specific information $m_l$ (e.g., road condition), so that we re-write $\gamma^Tm$ as 
\begin{equation}\label{static_dynamic}
    \gamma^T m = \gamma_d^T m_d + \gamma_l^T m_l,
\end{equation}
which decompose the contribution of $m$ into the sum of both terms. 

\begin{wrapfigure}[11]{r}{.3\textwidth}
    \centering
    \vspace{-0.1in}
    \includegraphics[width=\linewidth]{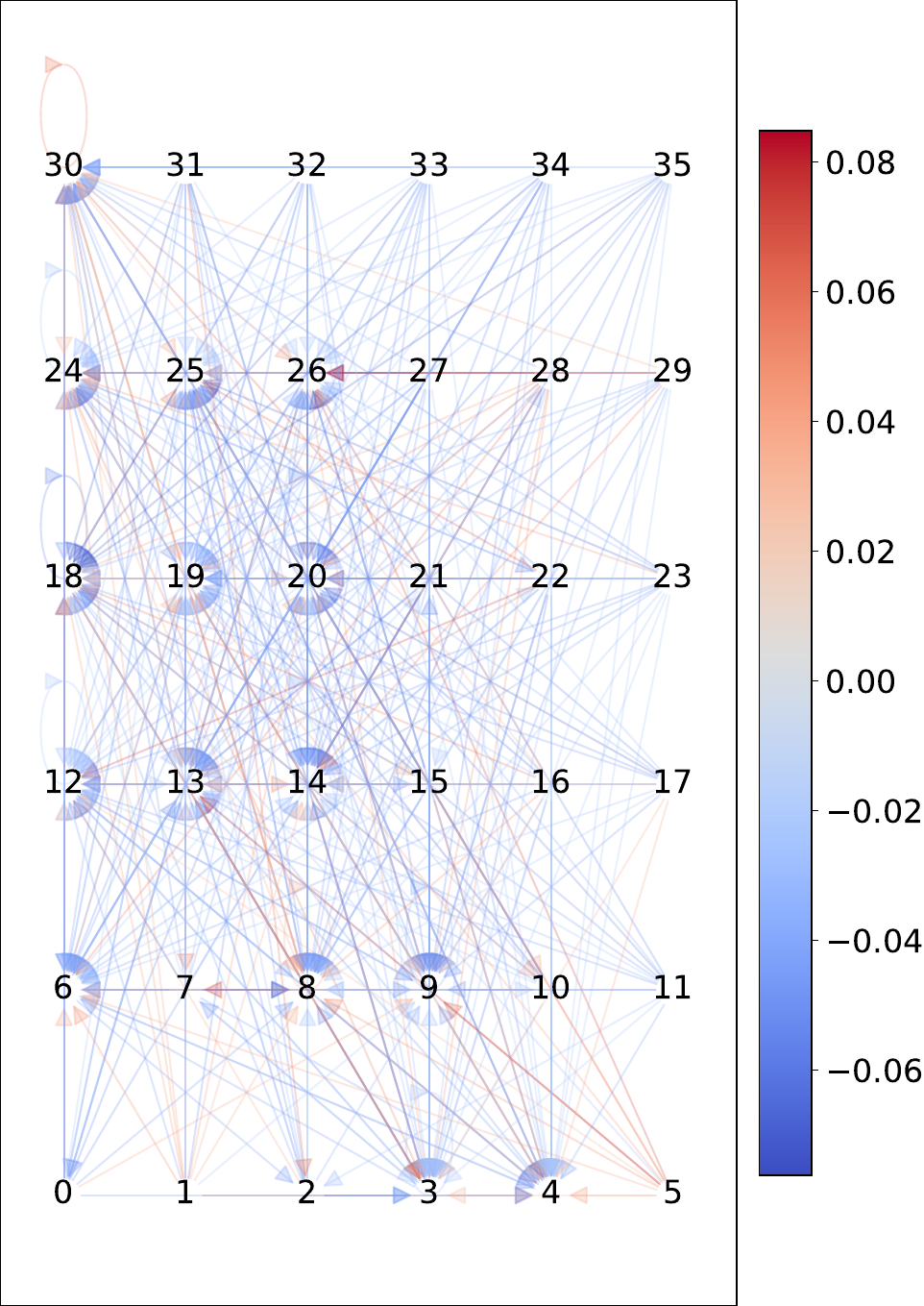}
    \caption{The distribution of $\alpha_{ij}$ closely follows the data distribution in Figure \ref{true_data_small}.}
    \label{fig:interaction_carved}
\end{wrapfigure}
Based on \eqref{static_dynamic}, we interpret the feature and interaction parameters of \LinearSTHawkes, estimated via Algorithm \ref{alter_algo}. First, Table \ref{tab:other_para} shows the estimated parameters for features (i.e., marks), whose magnitude indicates feature importance. Higher magnitude of estimates contribute more significantly to the growth of fire risk. Noticeably, the top two features in $\gamma_d$ (excluding summer, the seasonality parameter) are also factors in defining the \textit{Fire Danger Index}, which is a most commonly used index for fire hazard monitoring \cite{fdacsWildlandFire}. Therefore, the model estimates of feature parameters are physically meaningful. 
Next, Figure \ref{fig:interaction_carved} examines the location-to-location interaction parameters $\alpha_{ij}$, which is forced to be zero if centroids of two cells exceeds 4$\times$0.24 degrees. Values of $\alpha_{ij}$ above or below zero indicate excitatory or inhibitory effects from nearby and past events. The distribution of interaction effects closely aligns with the 2014---2017 training data in Figure \ref{true_data_small}. \rev{For instance, we see clusters of fire incidents in 2014-2017 training data in Figure \ref{true_data_small} around location 20 and as a result, location 20 in Figure \ref{fig:interaction_carved} also interacts intensively with its nearby neighbors. Quantitatively, if we use $\alpha_{ij}$ to roughly measure the amount of influence of location $i$ on location $j$:
    \begin{itemize}
        \item The amount of positive influence into location 20 (i.e., $\sum_{j:\alpha_{j,20}>0} \alpha_{j,20}$) is 0.40.
        \item The amount of negative influence into location 20 (i.e., $\sum_{j:\alpha_{j,20}<0} \alpha_{j,20}$) is -0.30.
        \item The amount of positive influence from location 20 (i.e., $\sum_{j:\alpha_{20,j}>0} \alpha_{20,j}$) is 0.29.
        \item The amount of negative influence from location 20 (i.e., $\sum_{j:\alpha_{20,j}<0} \alpha_{20,j}$) is -1.44.
    \end{itemize}}

\begin{table}[!t]
 \resizebox{\textwidth}{!}{%
\begin{tabular}[b]{p{3cm}p{2cm}p{2cm}p{3.5cm}|p{5cm}p{3cm}p{3cm}}
          &     \multicolumn{3}{c}{{\large Three Largest Estimates}}&      \multicolumn{3}{c}{{\large Three Smallest Estimates}} \\
\\
\hline
$\gamma_l$ estimate     &   0.301 &        0.231 &              0.184 &                 0.046 &         0.024 &           0.008 \\
$\gamma_l$ feature name &    Fire Tier1 &   Fire Tier2 &         Fire Tier3 &  PHYS=Developed-Roads &  PHYS=Conifer &  PHYS=Developed \\
\hline
$\gamma_d$ estimate          &   0.57 &        0.472 &               0.46 &                 0.217 &         0.117 &            0.02 \\
$\gamma_d$ feature name      &  Summer &  Temperature &  Relative Humidity &             LFP &         Spring &                  Winter \\
\hline
\end{tabular}%
}
\cprotect\caption{Estimated parameters of static marks $\gamma_l$ and dynamic marks $\gamma_d$ defined in \eqref{static_dynamic}. ``PHYS='' indicates road type or existing vegetation type. A larger parameter estimate indicates more contribution of the feature to fire hazards. Note that \textit{Temperature} and \textit{Relative Humidity} in $\gamma_d$ also define the widely-used Fire Danger Index so that \LinearSTHawkes \ selects physically meaningful features.}
\label{tab:other_para}
\end{table}

In addition, we can perform \textit{counterfactual analyses} using the estimated parameters: suppose a decision-maker wants to know the increase in risk when an external condition changes from $A$ to $B$ (e.g., Fire tier zone shift, changes in vegetation types, etc.). Then, the change in risk at a certain location and time is $\Delta(A,B):=\lambda(t,k,B)-\lambda(t,k,A)$. Similar analyses can be performed for a change in location from $k$ to $k_1$. Such analyses can help one better study the effect of different factors on fire risks, making risk management more effective.

\begin{figure}[!t]
    \centering
    \begin{minipage}{\textwidth}
    \centering
     \captionsetup[subfigure]{justification=centering}
        \includegraphics[width=\linewidth]{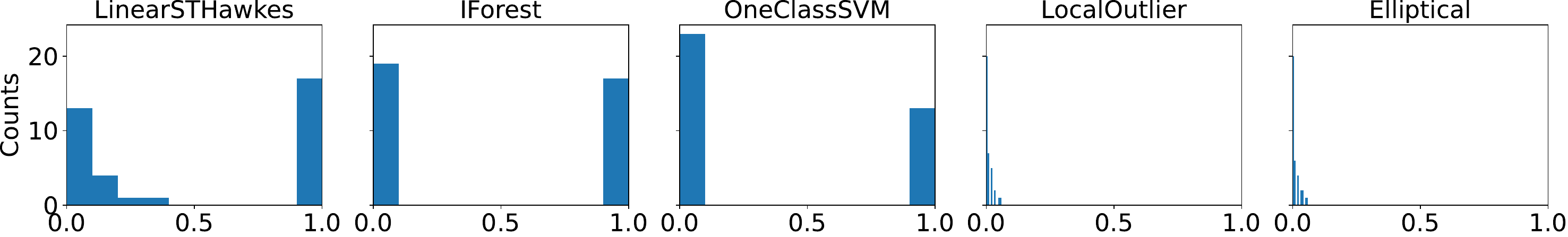}
        \subcaption{\texttt{LinearSTHawkes} $F_1$ scores against baselines}
        \label{carved_F1_hist}
    \end{minipage}

    \vspace{0.1in}
    \begin{minipage}{0.32\textwidth}
    \centering
     \captionsetup[subfigure]{justification=centering}
        \includegraphics[width=\linewidth]{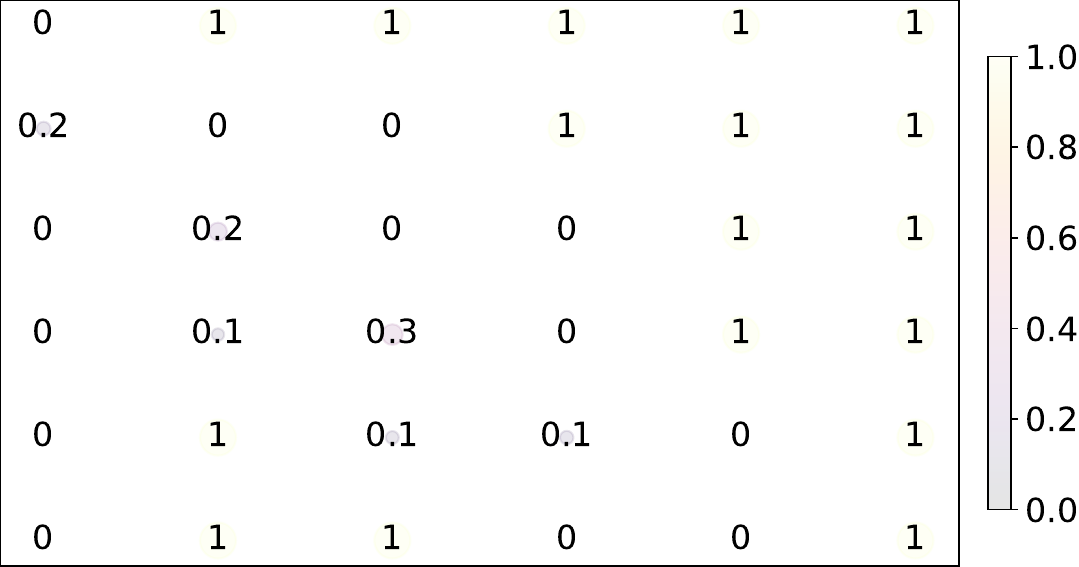}
        \subcaption{On map: $F_1$ score}
        \label{carved_f1_map}
    \end{minipage}
    \begin{minipage}{0.32\textwidth}
    \centering
     \captionsetup[subfigure]{justification=centering}
        \includegraphics[width=\linewidth]{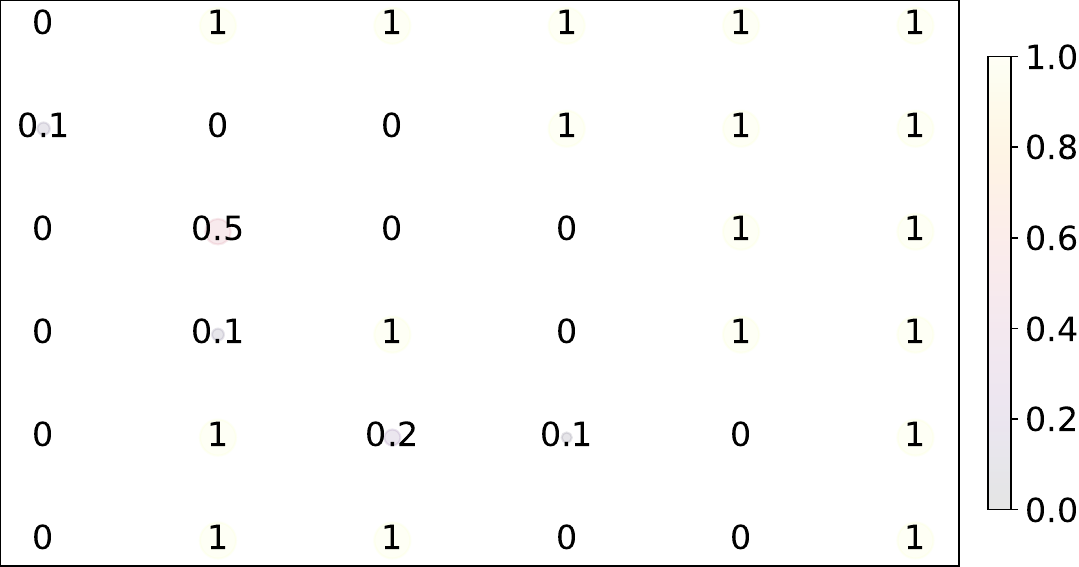}
        \subcaption{Recall}
    \end{minipage}
    \begin{minipage}{0.32\textwidth}
    \centering
     \captionsetup[subfigure]{justification=centering}
        \includegraphics[width=\linewidth]{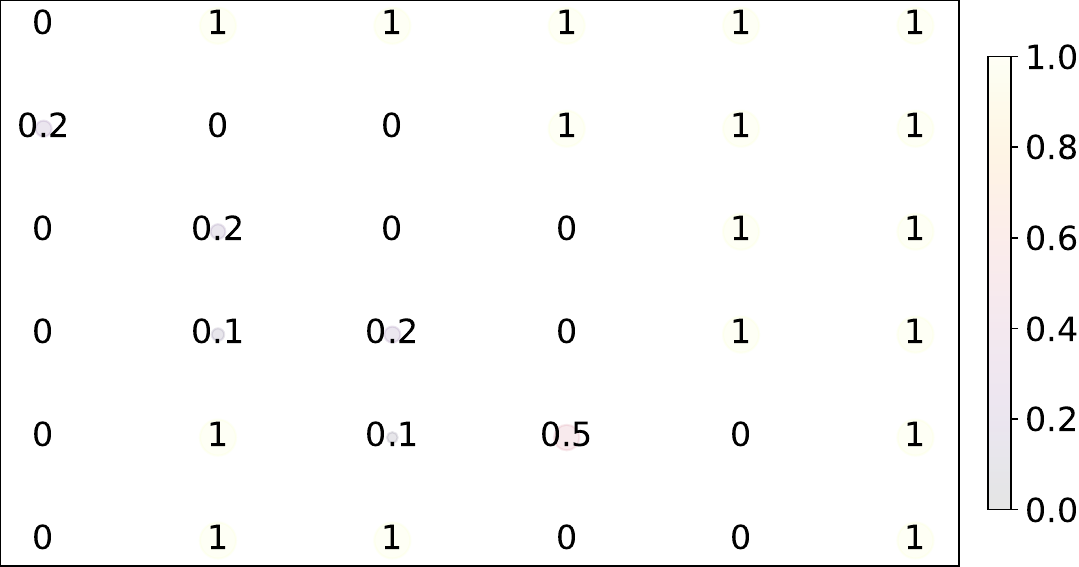}
        \subcaption{Precision}
        \label{carved_prec_map}
    \end{minipage}
    \cprotect\caption{Comparison across methods (top) and \LinearSTHawkes \ performance per location (bottom). Histograms of $F_1$ scores over all locations on the top row show that our \LinearSTHawkes \ outperforms other methods by yielding fewer zero $F_1$ scores, a moderate number of fractional $F_1$ scores, and more one $F_1$ scores. The bottom row visualizes the $F_1$ score, recall, and precision of \LinearSTHawkes \ at each location.}
\end{figure}

\vspace{0.1in}
\noindent \textit{Prediction results.} We first compare \LinearSTHawkes \ with several one-class classification baselines. We choose isolation forest \cite{iforest}, one-class SVM \cite{libsvm}, local outlier factor \cite{LOF}, and elliptic envelope \cite{Envelope} due to their popularity and generality. These classifiers, including static and dynamic marks, use the same data as \LinearSTHawkes. Figure \ref{carved_F1_hist} visualizes the histograms of $F_1$ scores by each method, which show that \LinearSTHawkes \ outperforms competing methods by yielding less zero $F_1$ scores and more one $F_1$ scores. Note that zero (resp. one) $F_1$ scores appear at locations that are the easiest (resp. hardest) to predict discussed earlier. In addition, \LinearSTHawkes \ can yield non-trivial fractional $F_1$ scores at other locations by capturing a decent number of true positives. Nevertheless, our model also yields many zero $F_1$ scores because the task is inherently challenging: it makes 365 daily predictions at each of 36 locations, in a total of 13140 predictions, when there are only 70 actual fire occurrences across all 36 locations.

\begin{figure}[!b]
    \centering
        \includegraphics[width=0.8\linewidth]{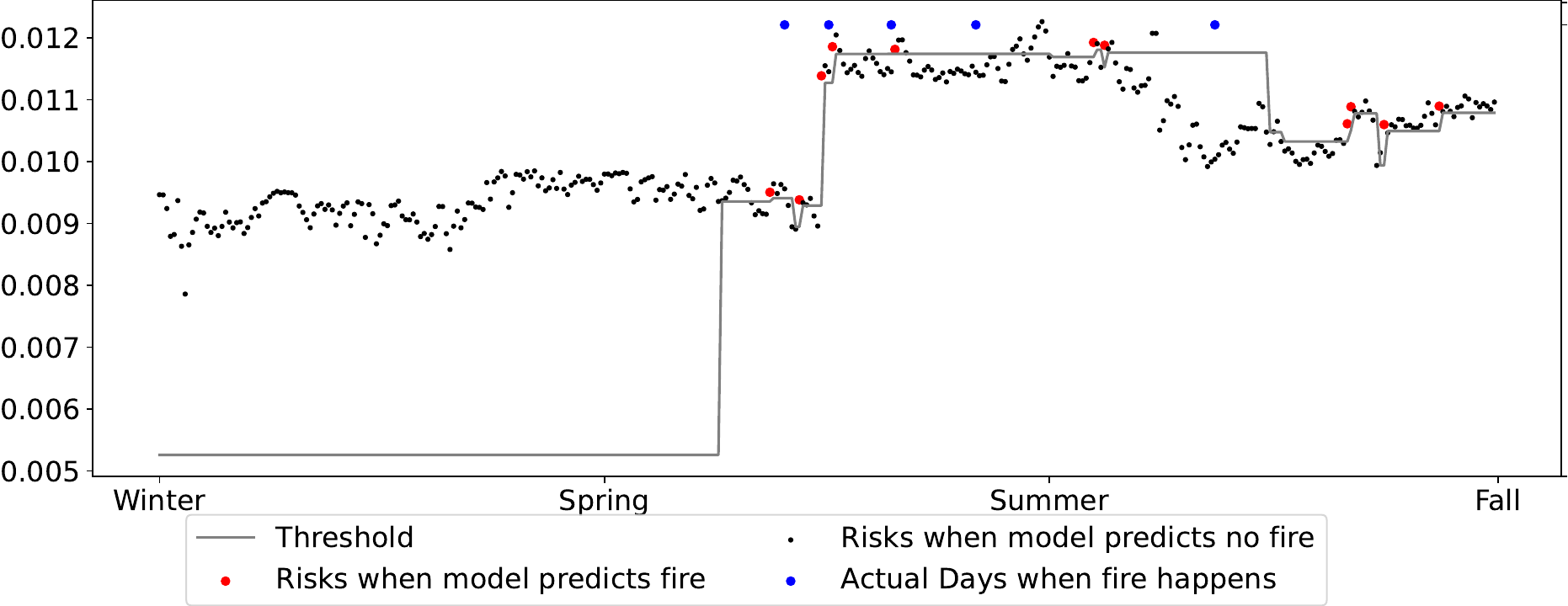}
    \cprotect \caption{Real-time prediction of fire risks and incidents on top of actual incidents and dynamic thresholds. The prediction by \LinearSTHawkes \ can closely match the actual data.}
    \label{fig:loc_pred_carved}
\end{figure}

We now illustrate the location-wise prediction results of \LinearSTHawkes. Figure \ref{carved_f1_map}---\ref{carved_prec_map} visualizes $F_1$ score, recall, and precision on each of the 36 location. The result helps us assess the prediction difficulty at various locations, where we suspect the difficulty arises partially due to the distribution shift \rev{of data in 2018 comparing to data in 2014-17} (cf. Figure \ref{true_data_small}). To better illustrate how \LinearSTHawkes \ makes a prediction, we further visualize in 
Figure \ref{fig:loc_pred_carved} the trajectory of rate prediction on top of actual incidents. Dynamic thresholds are obtained by using Algorithm \ref{algo:thres}. The figure shows that sharp increases in predicted fire risks tend to occur near true fire events, which helps us make correct predictions. In the future, to reduce the number of false positives, we may refit the model parameters during validation using newly observed incidents.

\subsection{Compare LinearSTHawkes vs. NonLinearSTHawkes}\label{sec:marked_vs_accum}

We now compare \LinearSTHawkes \ and \NonLinearSTHawkes \ on 2019 test data (cf. Figure \ref{true_data_small} right), where we train the feature extractor \rev{$g(m|t,k)$ in \eqref{eq:feature_extractor}} using the one-class SVM. In principle, one can use any feature extractor, but we choose SVM due to the flexibility of the kernel function. Based on earlier results, we only include seasonal and weather information, LFP, and FPI in the dynamic marks.

\begin{figure}[!t]
\begin{center}
    \begin{minipage}{0.33\textwidth}
    \centering
     \captionsetup[subfigure]{justification=centering}
        \includegraphics[width=\linewidth]{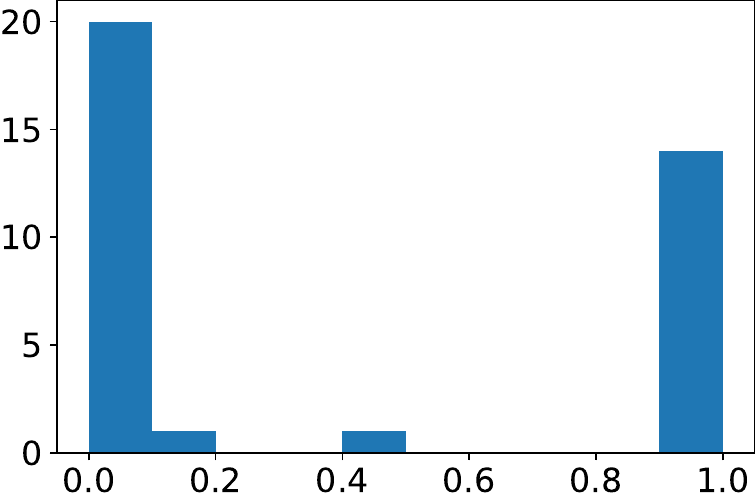}
        \subcaption{\LinearSTHawkes: $F_1$}
        \label{carved_F1compare_linear}
    \end{minipage}
    \begin{minipage}{0.63\textwidth}
    \centering
     \captionsetup[subfigure]{justification=centering}
        \includegraphics[width=\linewidth]{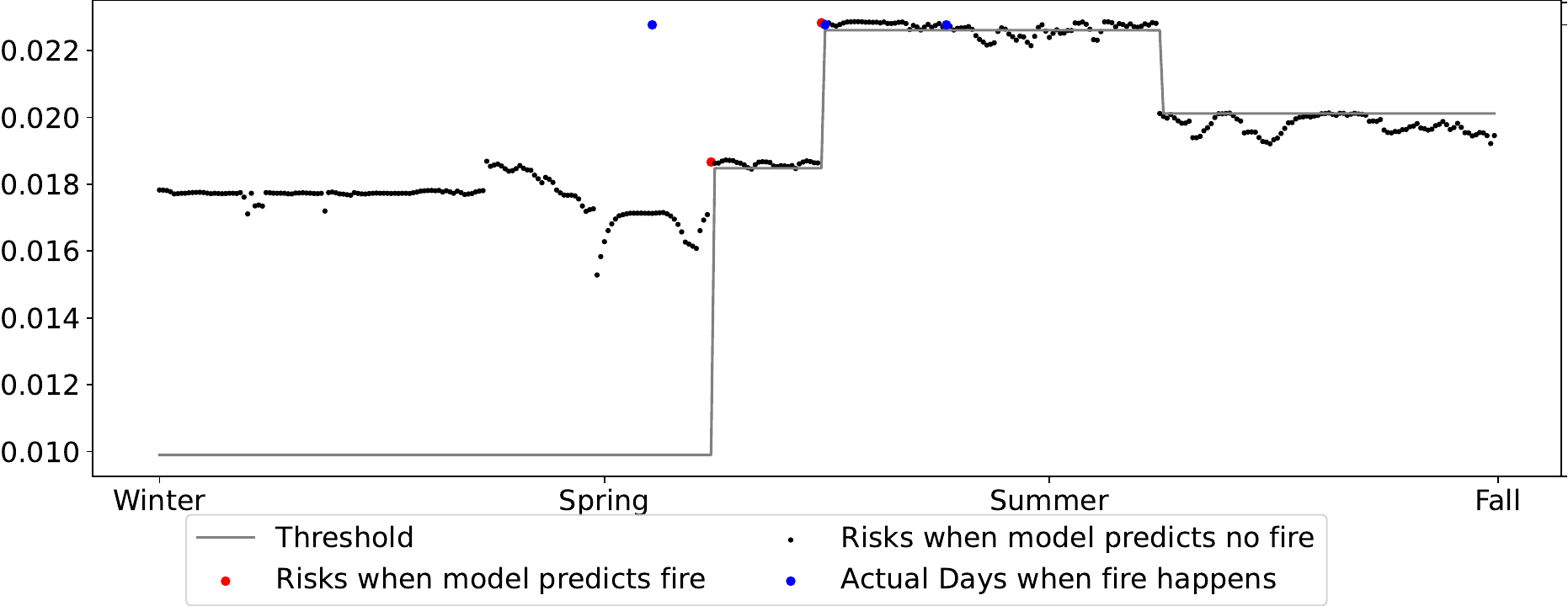}
    \end{minipage}
    \begin{minipage}{0.33\textwidth}
    \centering
     \captionsetup[subfigure]{justification=centering}
        \includegraphics[width=\linewidth]{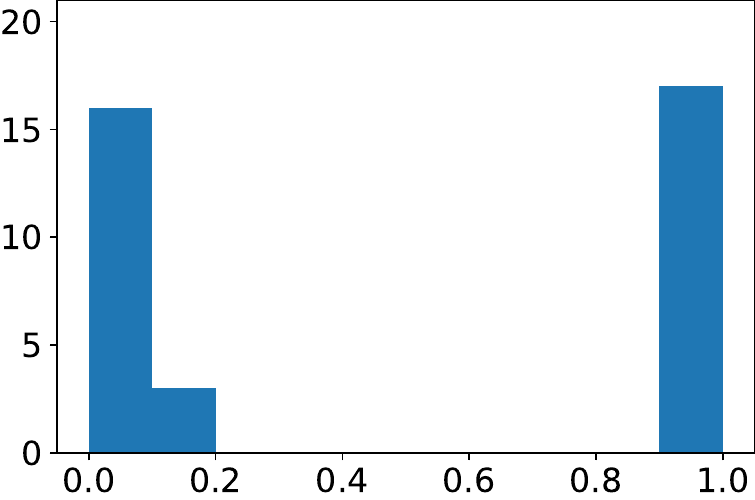}
        \subcaption{\NonLinearSTHawkes: $F_1$}
        \label{carved_F1compare_nonlinear}
    \end{minipage}
    \begin{minipage}{0.63\textwidth}
    \centering
     \captionsetup[subfigure]{justification=centering}
        \includegraphics[width=\linewidth]{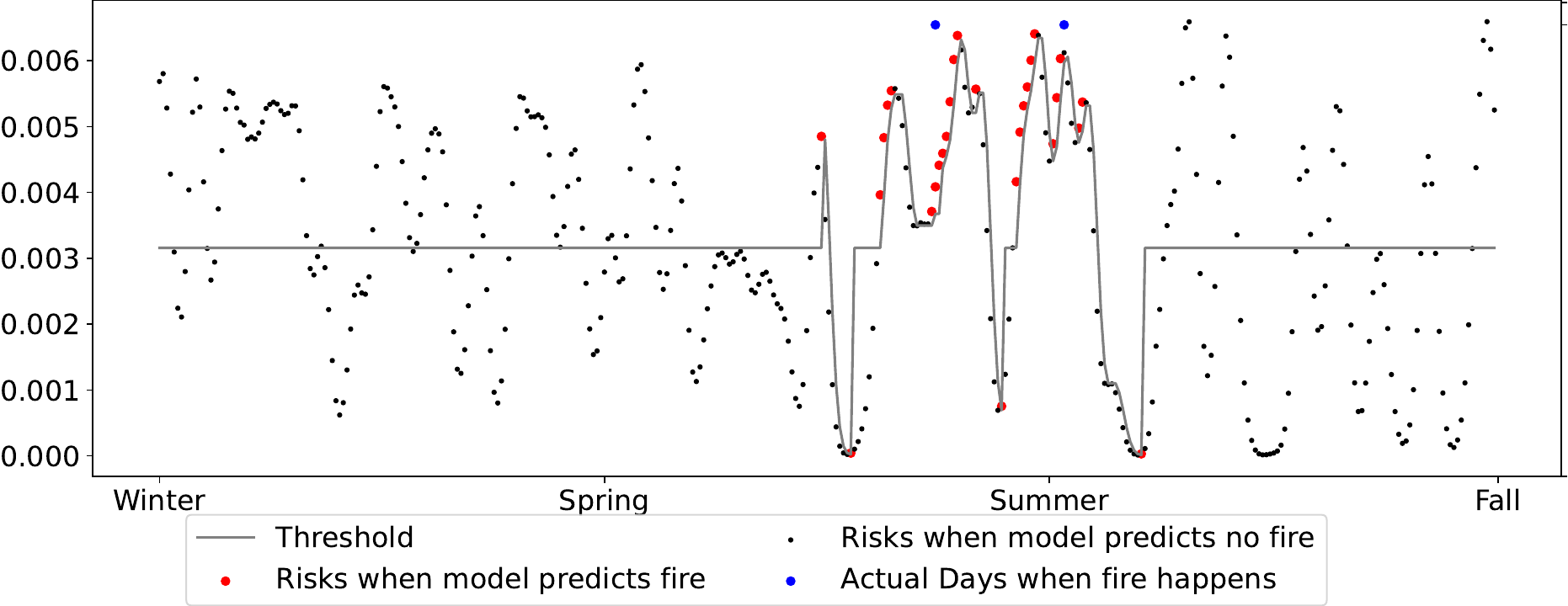}
    \end{minipage}
    \cprotect\caption{Compare \LinearSTHawkes \ with \NonLinearSTHawkes \ on 2019 test data. Both models are trained on 2014-2018 data. The top row shows results under \LinearSTHawkes, and the bottom row shows those under \NonLinearSTHawkes. In comparison, \NonLinearSTHawkes \ shows improved performance because of a more flexible feature extractor and the ability to yield less zero $F_1$ scores.}
    \label{fig:marked_vs_ImplicitSTHawkes}
    
    \end{center}
\end{figure}

Figure \ref{fig:marked_vs_ImplicitSTHawkes} compares the performance of both methods and there are several observations. First, the histograms of $F_1$ scores (cf. Figure \ref{carved_F1compare_linear} \& \ref{carved_F1compare_nonlinear}) show that \NonLinearSTHawkes \ performs better than \LinearSTHawkes, as the former yields more non-zero $F_1$ scores. To explain the improvement, we found the \rev{empirical} distribution of estimates \rev{$g(m|t,k)$} by \NonLinearSTHawkes{} \rev{to closely match} the Frechet distribution, a classic example from \textit{extreme value theory} \cite{de2006extreme}. \rev{Although the Frechet distribution is not used to aid modeling,} the connection allows \NonLinearSTHawkes{} to make a more accurate prediction because many rare events (e.g., fire incidents) follow the Frechet distribution. Further discussions appear in appendix \ref{appendix:EVT}. 
Second, the trajectory of predicted fire risks by \NonLinearSTHawkes{} (cf. Figure \ref{fig:marked_vs_ImplicitSTHawkes}, lower right) fluctuates much more than \LinearSTHawkes{} (cf. Figure \ref{fig:marked_vs_ImplicitSTHawkes}, top right). For this prediction task, such fluctuation enables better detection because actual fire incidents are often associated with sudden risk increases.

\begin{remark}[History-dependent mark in \NonLinearSTHawkes]\label{remark:mark}
Accumulated weather conditions can often induce fire events (e.g., several dry days earlier can lead to elevated fire risks). Thus, it seems natural to include in each $m_i$ additional spatio-temporal marks to account for accumulation effects. However, doing so has two drawbacks:
\begin{enumerate}[noitemsep,topsep=0em]
    \item Data acquisition and storage are much more expensive. One must collect a complete record of historical marks at each grid to fit the models. The issue mainly arises when the number of grids is large (e.g., hundreds) and marks frequently arrive (e.g., hourly). 
    \item The curse of dimensionality rises when each mark contains longer historical values. Note that the total number of fire incidents is fixed and typically small (e.g., hundreds over multiple years). Therefore, parameter estimation can be more difficult as the feature dimension increases. How to choose historical values appropriately to reduce the effect of this issue would increase difficulty in training.
\end{enumerate}
\end{remark}

\section{Large-Scale Data Validation}\label{sec:exp_hawkes}

\begin{figure}[!t]
\vspace{-0.15in}
\begin{center}
\begin{minipage}[t]{0.28\textwidth}
\centering\captionsetup[subfigure]{justification=centering}
\includegraphics[width=\linewidth]{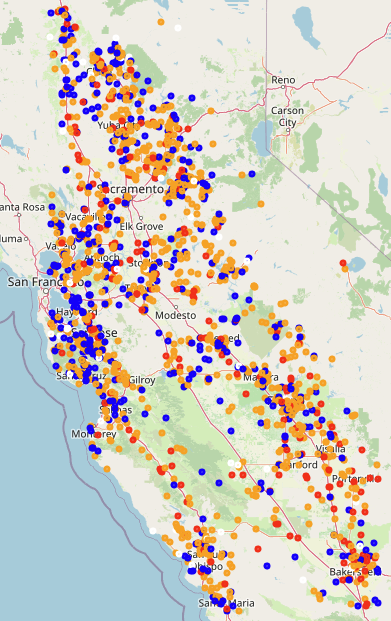}
\subcaption{Raw data}
\label{fig:data_vis1}
\end{minipage}
\hspace{0.1cm}%
\begin{minipage}[t]{0.298\textwidth}
\centering\captionsetup[subfigure]{justification=centering}
\includegraphics[width=\linewidth]{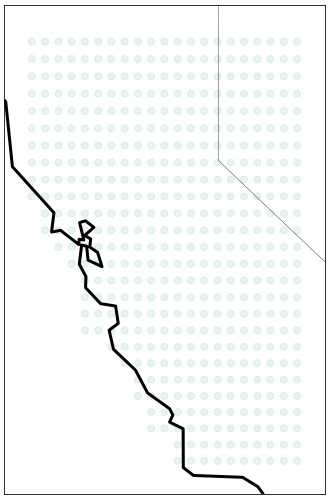}
\subcaption{Grid discretization}
\label{fig:grid_div}
\end{minipage}
\hspace{0.1cm}%
\begin{minipage}[t]{0.315\textwidth}
\includegraphics[width=\linewidth]{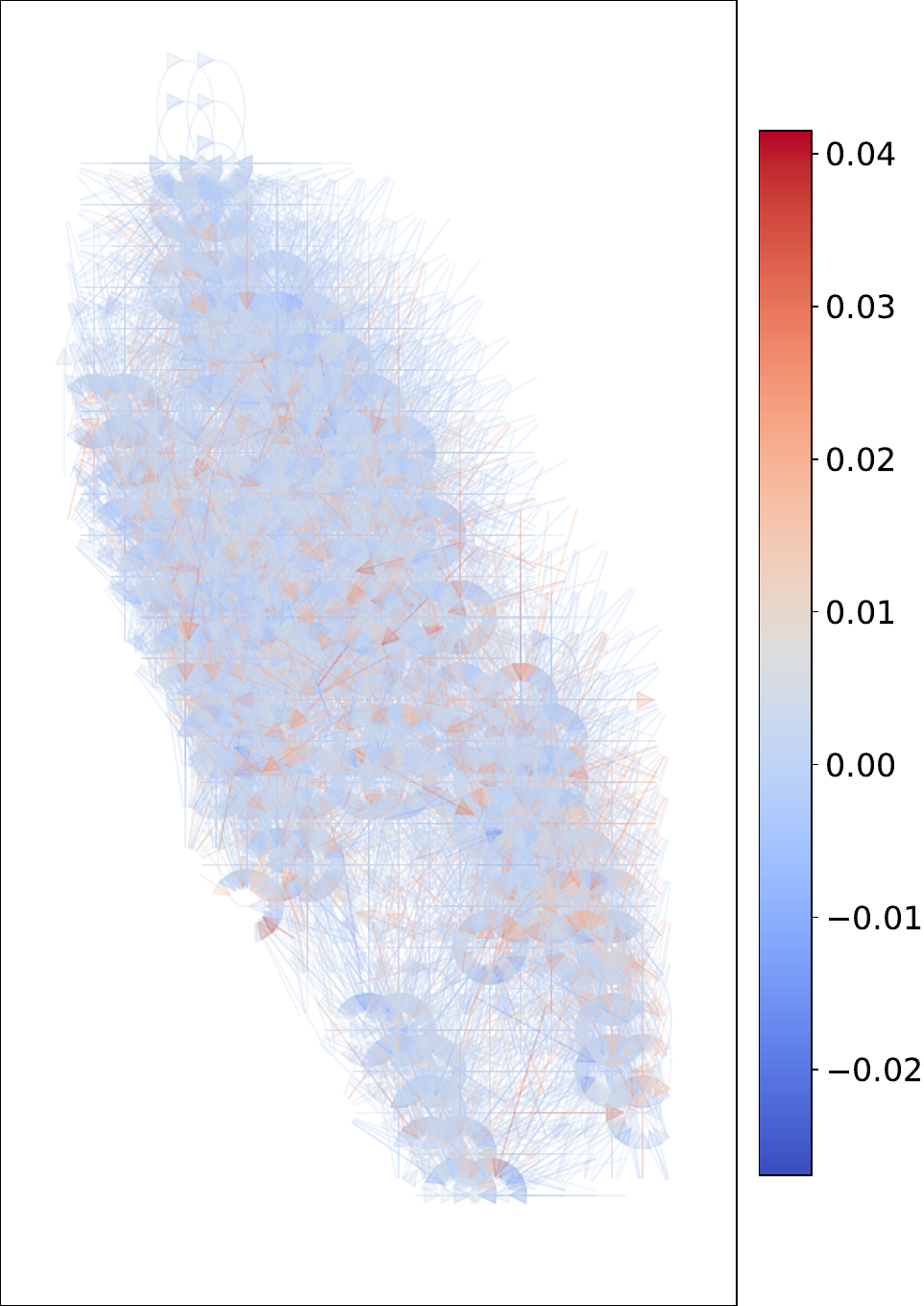}
\subcaption{Interaction $\alpha_{ij}$}
\label{fig:interaction}
\end{minipage}
\end{center}
\caption{Data visualization. (a) shows fire events colored by season as in Figure \ref{true_data_small}, (b) shows the grid discretization, and (c) visualizes the location-location interaction matrix parameters $\alpha_{ij}$.}
\end{figure}

We now show that our \LinearSTHawkes \ and \NonLinearSTHawkes \ are scalable to a large region with much more fire incidents and locations. There are a total of 2011 fire occurrences in this region, comprising 63\% of total wildfire incidents in California from 2014 to 2019. Figure \ref{fig:data_vis1} visualizes fire incidents within the region on the map, and Figure \ref{fig:grid_div} illustrates the resulting 453 grids after discretization into squares 
with side lengths equal to 0.24 degrees; we remove regions that lie inside the ocean. Most grids have no fire in the 5-year horizon since fire incidents seem to cluster near the coastal line with large populations. We remark that the setup and hyperparameter choices are the same as those in Section \ref{sec:exp_Hawkes_carved}. The distribution of estimated interaction parameters $\alpha_{ij}$ (cf. Figure \ref{fig:interaction}) still closely align with that of the actual data. \rev{For instance, Figure \ref{fig:data_vis1} shows there are clusters of true fire incidents around the coastal line on the west side and few incidents in the mid-south side. As a result, estimates in Figure \ref{fig:interaction} are much denser in distribution around the west side than around the mid-south side. As a concrete example, location 140 is on the west side along the coastal line, where there are clusters of fire incidents. Quantitatively, if we use $\alpha_{ij}$ to roughly measure the amount of influence of location $i$ on location $j$:
    \begin{itemize}
        \item The amount of positive influence into location 140 (i.e., $\sum_{j:\alpha_{j,140}>0} \alpha_{j,140}$) is 0.17.
        \item The amount of negative influence into location 140 (i.e., $\sum_{j:\alpha_{140}<0} \alpha_{j,140}$) is -0.30.
        \item The amount of positive influence from location 140 (i.e., $\sum_{j:\alpha_{140,j}>0} \alpha_{140,j}$) is 0.23.
        \item The amount of negative influence from location 140 (i.e., $\sum_{j:\alpha_{140,j}<0} \alpha_{140,j}$) is -0.47.
    \end{itemize}
    In comparison, location 20 is in the mid-south region of few clusters of fire incindents. Quantitatively, if we use $\alpha_{ij}$ to roughly measure the total influence of location $i$ on location $j$:
    \begin{itemize}
        \item The amount of positive influence into location 20 (i.e., $\sum_{j:\alpha_{j,20}>0} \alpha_{j,20}$) is 0.00.
        \item The amount of negative influence into location 20 (i.e., $\sum_{j:\alpha_{j,20}<0} \alpha_{j,20}$) is -0.09.
        \item The amount of positive influence from location 20 (i.e., $\sum_{j:\alpha_{20,j}>0} \alpha_{20,j}$) is 0.00.
        \item The amount of negative influence from location 20 (i.e., $\sum_{j:\alpha_{20,j}<0} \alpha_{20,j}$) is 0.00.
    \end{itemize}}

\subsection{Real-time fire risk prediction}

\begin{figure}[!t]
    \centering
    \begin{minipage}{\textwidth}
        \centering
        \captionsetup[subfigure]{justification=centering}
        \includegraphics[width=\linewidth]{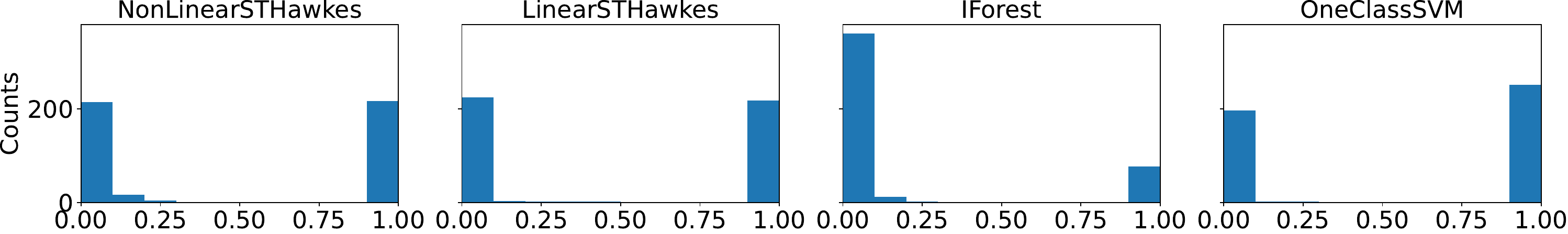}
        \cprotect \subcaption{\LinearSTHawkes \ and \NonLinearSTHawkes \ $F_1$ score comparison with baselines}
        \label{large_histo}
    \end{minipage}
    
    \vspace{0.1in}
     \begin{minipage}{\textwidth}
        \centering
        \captionsetup[subfigure]{justification=centering}
        \includegraphics[width=\linewidth]{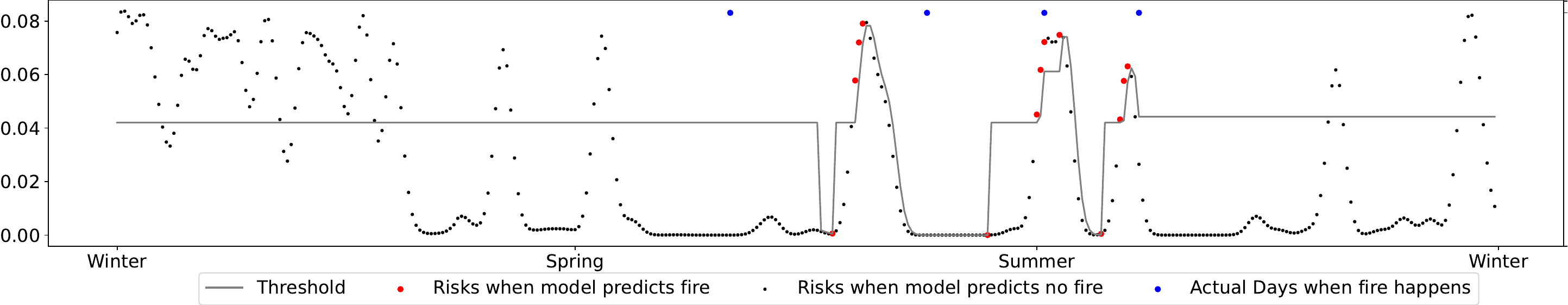}
        \subcaption{\NonLinearSTHawkes \ real-time prediction}
        \label{large_pointwise}
    \end{minipage}
    \cprotect\caption{On 2019 test data: \rev{The top row} compares the histograms of $F_1$ score under various methods. The leftmost \NonLinearSTHawkes \ has the most number of non-zero $F_1$ scores, with many being 1. \rev{The bottom row} visualizes the temporal  predicted risks by \NonLinearSTHawkes \ at one grid. Overall, \NonLinearSTHawkes \ yields the best performance among all models. }
    \label{fig:F1_comparison}
\end{figure}

Figure \ref{large_histo} compares the prediction performances of \NonLinearSTHawkes, \LinearSTHawkes, IForest, and OneClassSVM. We see that \NonLinearSTHawkes \ performs better than both the \LinearSTHawkes \ and the isolation forest by yielding more non-zero $F_1$ scores and a large number of $F_1$ scores being one. Due to its flexible feature extractor, the \NonLinearSTHawkes \ is also competitive against the one-class SVM; importantly, it yields more $F_1$ scores between zero and one, making it more informative than the one-class SVM on certain locations. Hence, \NonLinearSTHawkes \ maintains improved performance than other models even if the number of grids significantly increases. Figure \ref{large_pointwise} further visualizes the real-time prediction behavior of \NonLinearSTHawkes, where the peaks identified as fire incidents closely align with the actual incidents.

\subsection{Fire magnitude conformal prediction sets}\label{sec:exp_UQ}

\begin{figure}[!t]

   \begin{minipage}{\textwidth}
          \centering \includegraphics[width=.7\linewidth]{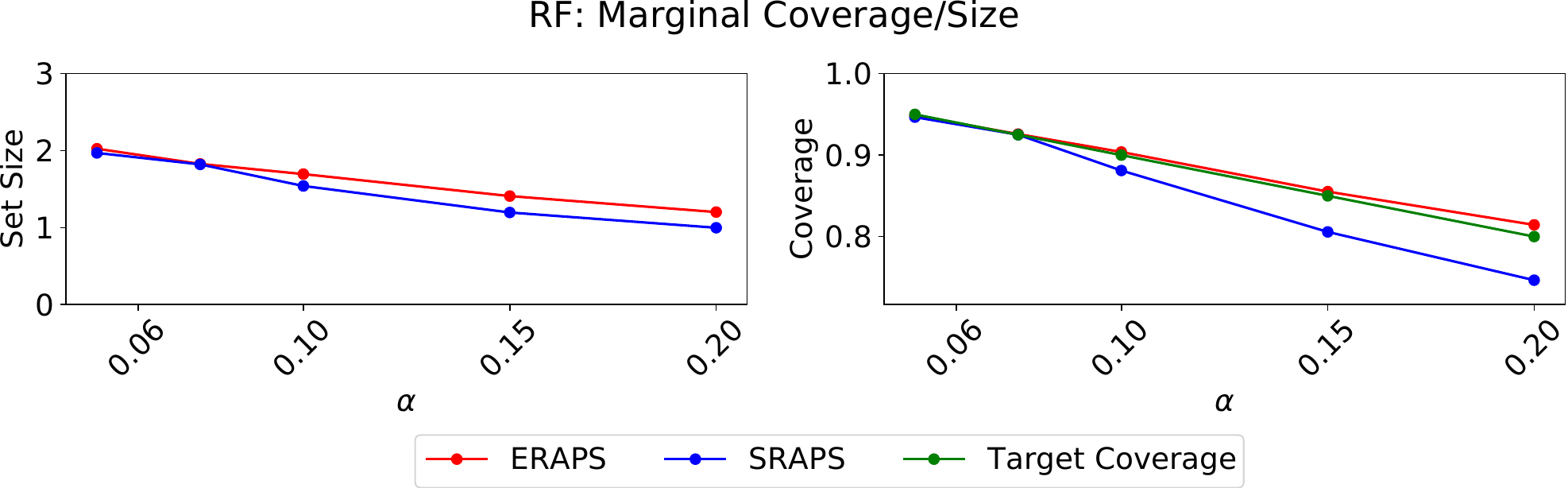}
   \end{minipage}
   
   \vspace{0.1in}
   \begin{minipage}{\textwidth}
       \centering
       \includegraphics[width=.7\linewidth]{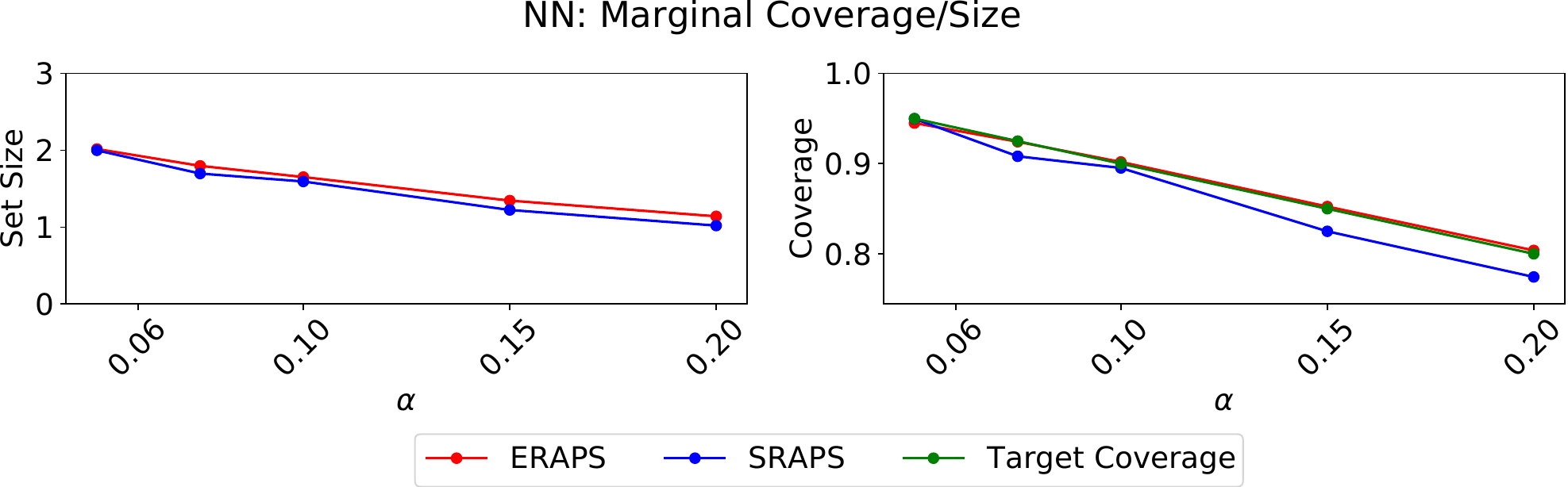}
   \end{minipage}
    
    \cprotect \caption{Marginal coverage \eqref{coverage} and size of prediction sets by \ERAPS \ and \SRAPS \ under the random forest classifier and the neural network classifier. \ERAPS \ always maintains desired coverage, whereas competing methods can fail to do so.}
    \label{fig:ERAPS_marginal_RF}
\end{figure}

We show that prediction sets by \ERAPS \ maintain desired coverage defined in \eqref{coverage}. Data in 2014-2018 are training data, and data in 2019 are test data, where there are a total of five possible fire magnitude. Both the random forest classifier (RF) and the neural network classifier (NN) are used as prediction algorithms; their setup is the same as those in \cite{xuERPAS2022}. \rev{We let regularization parameters $(\lambda, k_{\text{reg}})=(1,2)$ as suggested in \cite{xuERPAS2022}.} Figure \ref{fig:ERAPS_marginal_RF} shows marginal coverage under both classifiers, where we also compare \ERAPS \ against a competing method \rev{titled \textit{split regularized adaptive prediction set}} (\SRAPS{}) \cite{MJ_classification}. The details of \SRAPS \ are described in \cite[Algorithm 1]{xuERPAS2022}. We have two findings. First, \ERAPS \ performs very similarly under both classifiers and always maintains $1-\alpha$ coverage, whereas \SRAPS \ tends to lose coverage at different values of $\alpha$. Thus, \ERAPS \ is more robust and consistent in terms of coverage. Second, both methods return prediction sets with almost the same sizes, but \ERAPS \ is preferable due to its ability to maintain near $1-\alpha$ coverage. 

\section{Conclusion and Discussions}\label{sec:conclude}

We have developed a predictive framework for wildfire risk and magnitude using multi-modal sensing data, based on a mutually exciting spatio-temporal point process model as well as time series \rev{CP} set. We established performance guarantees of the proposed methods, and demonstrate the good performance on large-scale real data experiments. Overall, our method is efficient in model parameter, enjoys interpretability,  accurate prediction against existing methods. 
There are several future works. Regarding the point process model, we can consider beyond the parametric forms in \eqref{param_f} and \eqref{eq:param_marks}, such as the more general neural network-based formulations. The development of dynamic marks in Algorithm \ref{algo:thres} can also be refined. Regarding conformal uncertainty quantification, remaining questions include how to better utilize the existing time-series method when data have an additional spatial dimension. 

\rev{From our numerical results, we observe that distribution shifts may exist sometime for wildfire prediction. 
Although our \LinearSTHawkes{} and \NonLinearSTHawkes{} are not designed to explicitly consider distribution shift, they still yield improved performance against baseline models on real data. In particular, as shown in Fig. \ref{carved_F1_hist} on small-scale data and Fig. \ref{fig:F1_comparison} on large-scale data, our proposed models always outperform the baseline one-class classifiers. As a result, although the performance of our proposed framework may vary from year to year, it is still preferable in terms of predictive ability. We believe this is due to the model design to capture spatial-temporal information (e.g., past fire incidents around neighbors) and mark contribution (e.g., how multi-modal sensor information contributes to fire risks). To mitigate the adverse effects of distribution shifts, one approach is to introduce uncertainty into model parameters. For instance, instead of specifying the parameters in the optimization problem \eqref{mod} as unknown constants in our models, one could allow them to vary within a pre-specified range (or even treat them as random variables). With accurate parameter estimation, the estimated model could better address model shifts that arise from distribution shifts in test data. However, we do not explore this model design in this work, as our goal is to propose simple yet effective models for capturing fire risks using multi-modal data and establishing theoretical guarantees based on the proposed models (see Theorem \ref{theory_param_recovery}).}

\section*{Acknowledgement}

This work is partially supported by an NSF CAREER CCF-1650913, and NSF DMS-2134037, CMMI-2015787, DMS-1938106, and DMS-1830210, and in part by the U.S. Department of Energy Advanced Grid Modeling
Program under Grant DE-OE0000875.

\bibliographystyle{IEEEtran} 
\bibliography{ref}

\appendices

\section{Proof}\label{sec:proof}
\subsection{Proof of Lemma \ref{para_recovery_lem}}\label{proof_lem1}
Under the projected gradient descent \eqref{pgd}, we have
\begin{align*}
    \|\theta_k-\theta^*\|_2^2
    & = \|\text{Proj}_{\Theta}(\theta_{k-1}-t_k F(\theta_{k-1})-\theta^*)\|_2^2 \\ 
    & \leq \|\theta_{k-1}-t_k F(\theta_{k-1})-\theta^*\|_2^2 \\
    & = \|\theta_{k-1}-\theta^*\|_2^2 - 2t_kF(\theta_{k-1})^T[\theta_{k-1}-\theta^*] + t_k^2\|F(\theta_{k-1})\|_2^2.
\end{align*}
By \rev{assumptions} \eqref{strongly_monotone} and \eqref{bounded_grad} on the monotone operator $F$ and the fact that $F(\theta^*)=0$ when $\theta^*$ is the minimizer of $f$, we have
\begin{equation}\label{D_k_prelim}
     \|\theta_k-\theta^*\|_2^2 \leq (1-2t_k\kappa)\|\theta_{k-1}-\theta^*\|_2^2+t_k^2M^2.
\end{equation}
Define $\rev{d_k}:=\|\theta_k-\theta^*\|_2^2$. If $S:=M^2/\kappa^2$ and $t_k=[\kappa(k+1)]^{-1}$, we show by induction that 
\begin{equation}\label{small_Dk}
    \rev{d_k} \leq \frac{S}{k+1} = \frac{M^2}{\kappa^2(k+1)}.
\end{equation}
\textbf{Base case $k=0$.} Pick $\theta, 
\theta'$ such that $\|\theta-\theta'\|_2=D$, 
\rev{where $D$ in \eqref{bonded_domain} denotes the diameter of the parameter set for $\theta$.} Observe that 
\begin{align*}
    MD & \geq [F(\theta)-F(\theta')]^T[\theta-\theta']\\
        & \geq \kappa \|\theta-\theta'\|^2_2 = \kappa D^2.
\end{align*}
Thus, $D\leq 2M/\kappa$. By \rev{assumption} \eqref{bonded_domain}, we \rev{thus have} that $\rev{\sqrt{d_0}}=\|\theta_0-\theta^*\|_2\leq D$, so that the base case is proven.

\noindent \textbf{Induction step from $k-1$ to $k$, $k\geq 1$.} Observe that by the choice of $t_k$, $\kappa t_k=(k+1)^{-1}\leq 1/2$. Thus
\begin{align*}
    \rev{d_k} 
    & \leq (1-2t_k\kappa)\rev{d_{k-1}}+t_k^2M^2 \qquad \text{[By \eqref{D_k_prelim}]}\\ 
    & \leq  (1-2t_k\kappa)\frac{S}{k} + t_k^2M^2  \quad 
 \ \qquad \text{[By induction hypothesis and } \kappa t_k\leq 1/2] \\ 
    & = (1-\frac{2}{k+1})\frac{S}{k}+\frac{S}{(k+1)^2} = (\frac{k-1}{k}+\frac{1}{k+1})\frac{S}{k+1} \leq \frac{S}{k+1}.
\end{align*}

\subsection{Proof of Theorem \ref{para_recovery_final}}\label{proof_thm1}
First, note that after searching over $J$ grid points of $\beta_j$ in the region $[\beta_0,\beta_1]$, we obtain
    \begin{equation}\label{beta_bound}
        \|\beta_{j^*}-\beta^*\|^2_2 \leq \frac{\beta_1-\beta_0}{J^2}.
    \end{equation}
    Meanwhile, we know that for each fixed value of $\beta_j$, the function $\ell(\beta_j,\theta[\beta_j])$ is convex in $\theta[\beta_j]$. Because the constrains when solving for \eqref{loglik} are also convex, Lemma \eqref{para_recovery_lem} implies
    \begin{equation}\label{theta_beta_bound}
        \|\hat{\theta}[\beta_{j^*}]-\theta^*[\beta_{j^*}]\|^2_2 = \mathcal{O}((k+1)^{-1})
    \end{equation}
    after $k$ projected gradient descent steps \eqref{pgd}. Putting \eqref{beta_bound} and \eqref{theta_beta_bound} together, we thus have 
    \[
    \|\hat{\theta}-\theta^*\|_2^2 = \mathcal{O}\left(\frac{1}{J^2}\right)+\mathcal{O}\left(\frac{1}{k+1}\right)= \mathcal{O}\left(\frac{1}{J^2}+\frac{1}{k+1}\right)
    \]

\subsection{Proof of Lemma \ref{lem:tildeFandhatF}}\label{proof_lem2}
The proof is identical to that of \cite[Lemma 2]{EnbPI} so we omit the mathematical details. The gist of the proof proceeds by bounding the size of the set of past $N$ estimated non-conformity scores which deviate too much from the oracle one. The set is denoted as
\[
S_N:=\{i \in [N]: |\hat \tau_i-\tau_i|>\vartheta_N^{2/3} \}.
\]
Then, one can relate the difference $|\tildeF{x}-\hatF{x}|$ at each $x$ to a sum of two terms of indicator variables--ones whose index belongs to $S$ and ones which does not. The ones that does not belong to $S$ can be bounded using the term $|\tildeF{x}-F(x)|$ up to a multiplicative constant.

\subsection{Proof of Lemma \ref{lem:tildeFandF}}\label{proof_lem3}
The proof is identical to that of \cite[Lemma 1]{EnbPI} so we omit the mathematical details. In fact, this is a simple corollary of the famous Dvoretzky–Kiefer–Wolfowitz
inequality \cite[p.210]{Kosorok2008IntroductionTE}, which states the convergence of the empirical bridge to actual distributions under the i.i.d. assumption.

\subsection{Proof of Theorem \ref{thm:asym_cond_cov}}\label{proof_thm2}
The proof is identical to that of \cite[Theorem 1]{EnbPI} so we omit the mathematical details. The gist of the proof proceeds by bounding the non-coverage $|\PP(Y_i \notin \widehat{C}(X_i,\alpha))-\alpha|$ using the sum of constant multiples of $\sup_{x}|\tildeF{x}-\hatF{x}|$ and $\sup_{x}|\tildeF{x}-F(x)|$, both of which can be bounded by Lemmas \ref{lem:tildeFandhatF} and \ref{lem:tildeFandF} above.

\subsection{Proof of Theorem \ref{thm:asy_set}}\label{proof_thm3}
Based on the assumptions and the definition in \eqref{eq:set}, we now have
\begin{align*}
    C^*(X_i,\alpha)&=\{1,\ldots,c^*\}, c^*=\arg\max_{c} \tau_i(c) < F^{-1}(1-\alpha),\\
    \widehat{C}(X_i,\alpha)&=\{1,\ldots,\hat{c}\}, \hat{c}=\arg\max_{c} \hat{\tau}_i(c) < \hat{F}^{-1}(1-\alpha),
\end{align*}
where $\hat{F}^{-1}$ is the empirical CDF based on estimated non-conformity scores $\{\hat{\tau}_{i-N},\ldots,\hat{\tau}_{i-1}\}$.
We now show that $ \widehat{C}(X_i,\alpha) \Delta C^*(X_i,\alpha) \leq 1$ if and only if 
\[
\| \hat{\tau}_i- \tau_i\|_{\infty} \rightarrow 0 \text{ and }\hat{F}^{-1}(1-\alpha)\rightarrow F^{-1}(1-\alpha).
\]
($\Rightarrow$) Without loss of generality, suppose that $\hat{c}<c^*$ so that $ \widehat{C}(X_i,\alpha) \Delta C^*(X_i,\alpha) > 1$. Then, by definition of the prediction sets, we must have 
\begin{align*}
    & \hat{\tau}_i(c^*) \geq \hat{F}^{-1}(1-\alpha), \\
    & \tau_i(c^*) < F^{-1}(1-\alpha).
\end{align*}
Denote $\delta_{\tau,i}:=\hat{\tau}_i(c^*)-\tau_i(c^*)$ and $\delta_{F}:=F^{-1}(1-\alpha)-\hat{F}^{-1}(1-\alpha)$, we thus have 
\[
\delta_{\tau,i}+\delta_{F}\geq F^{-1}(1-\alpha)-\tau_i(c^*)>0.
\]
However, this is a contraction when $N$ approaches infinity---by the assumption that $\| \hat{\tau}_i- \tau_i\|_{\infty} \rightarrow 0$ and the earlier results that $\hat{F}^{-1}(1-\alpha)\rightarrow F^{-1}(1-\alpha)$, we must have $\delta_{\tau,i}$ and $\delta_{F}$ both converging to zero.

\noindent ($\Leftarrow$) By the form of the estimated and true prediction sets, it is obvious that if $\| \hat{\tau}_i- \tau_i\|_{\infty} \rightarrow 0$ and $\hat{F}^{-1}(1-\alpha)\rightarrow F^{-1}(1-\alpha)$, their set difference must converges to zero.

\section{Additional details}\label{appendix}
\subsection{Log-likelihood derivation}\label{log-lik-deriv}

The first two terms under $\log$ can be trivially derived upon substitution, so we only simplify the integration term: 
\begin{align*}
    \sum_{k=1}^K\int_0^T \lambda_g(\tau,k)d\tau 
    &= \sum_{k=1}^K\int_0^T (\mu(k)+\underset{j:t_j<t}{\sum}\alpha_{u_j,k}\beta e^{-\beta(\tau-t_j)})d\tau\\
    &= \sum_{k=1}^K T \mu(k) + \sum_{k=1}^K \sum_{j=1}^n \int_0^T \textbf{1}(\tau>t_j)\alpha_{u_j,k}\beta e^{-\beta(\tau-t_j)})d\tau\\
    &\overset{(i)}{=} \sum_{k=1}^K T \mu(k) + \sum_{k=1}^K \sum_{j=1}^n \alpha_{u_j,k}(1-e^{-\beta(T-t_j)}),
\end{align*}
where (i) follows from the definite interval formula for exponential functions. Interchanging the finite sums $\sum_{k=1}^K \sum_{j=1}^n$ yields (\ref{loglik}).

Under the general formulation \eqref{mod1}, we have $\lambda_g(t,k)=\mu(k)+\underset{j:t_j<t}{\sum} \mathcal K(u_j,k,t_j,t)$, so that the integral is simplified as
\begin{equation*}
    \sum_{k=1}^K T \mu(k) + \sum_{k=1}^K \sum_{j=1}^n \int_{t_j}^T \mathcal K(u_j,k,t_j,\tau) d\tau,
\end{equation*}
which may not have a closed form expression. In particular, there have been many parametric and non-parametric forms for $\lambda_g(t,k)$, including neural network-based models discussed in the literature review. Although they are more flexible and potentially more effective, the log-likelihood objective becomes non-convex, requiring gradient-descent type methods for local optimization under more computational resources.

\subsection{Alternating minimization}\label{appendix:altermin}
Denote $\theta[\beta]=\theta-\{\beta\}$ so that $\theta[\beta]$ contains all parameters except $\beta$ and $\theta[\beta] \cup \beta = \theta$. We then define 
\[\Psi(\theta[\beta],\beta):=-\ell(\theta).
\]
Algorithm \ref{alter_algo} 
contains details for the alternating minimization procedures. It first finds minimizers of $\Psi(\theta[\beta],\beta)$, given $\beta^0$ as the initial value of the one-dimensional parameter $\beta$. 
Then, we can use one-dimensional line search to solve for $\beta$, given the other estimates. 
The procedure iterates for a total of $N$ times, \rev{where we describe the computational efficiency of the proposed approach in Remark \ref{remark_computation}.} In general, we can allow $\beta$ to be location-dependent, such as having the same support as $\alpha_{u_i,k}$.

\begin{remark}[Parameters] \hfill
\begin{itemize}
\itemsep=0em
    \item $\beta^0$ is the initial guess of the temporal influence parameter, whose value depends on problem context. It can typically be set to 1.
    \item The lower end $\rev{\beta_{\text{low}}}$ (in line 3, Algorithm\ref{alter_algo}) can remain constant since we know $\beta>0$, so that a reasonably small $\rev{\beta_{\text{low}}}$ suffices.
    \item $\epsilon_{\beta}$ determines the stopping criteria, whose choice depends on the desired degree of accuracy.
\end{itemize}
\end{remark}

\begin{remark}[Algorithm Details]\hfill
\begin{itemize}
\itemsep=0em
    \item The termination criterion (Line 4-6, Algorithm \ref{alter_algo}) can be justified: once consecutive solutions for $\beta$ are close to each other, the solutions for $\theta[\beta]$ are likely to be close to each other in vector norm. 
    \item Since $\Psi (\theta[\beta],\beta)$ is non-convex in $\beta$, the one-dimensional line search is only guaranteed to find a local minimum. Nevertheless, once $\theta[\beta]^{(k)}$ is computed by Algorithm \ref{alter_algo}, we can clearly characterize the number of local minima of $\Psi (\theta[\beta]^{(k)},\beta)$. If it has multiple local minima within the bisection search domain, we can use line search multiple times to find the global minimum. Doing so is efficient because the search region for $\beta$ doubles every time (e.g., $K$ is logarithmic in widths of the search region) and evaluating the derivative of $\Psi (\theta[\beta]^{k},\beta)$ at each possible minimizer is a constant operation.
\end{itemize}
\end{remark}

\begin{algorithm}[htbp]
\caption{Alternating Minimization for Regularized Marked Spatio-Temporal Hawkes Process Model (Eq. (\ref{mod}))}
\label{alter_algo}
\begin{algorithmic}[1]
\Require{$\beta^0,K,\rev{\beta_{\text{low}}},\epsilon_{\beta}$}
\Ensure{$\theta[\beta]^*, \beta^*$}
\For {$k = 1, \dots, K$}
\State $\theta[\beta]^{(k)} \gets \underset{\theta[\beta]}{\arg\min} \Psi(\theta[\beta],\beta^{(k-1)})$ using convex optimization solvers (e.g., \Verb|CVX| \cite{CVX})
\State $\beta^{(k)} \gets \underset{\beta}{\arg\min} \Psi(\theta[\beta]^{(k)},\beta)$ using one-dimensional line-search (e.g., within $[\rev{\beta_{\text{low}}}, 2^k], \rev{\beta_{\text{low}}}>0$).
\If{$|\beta^{(k)}-\beta^{(k-1)}|\leq \epsilon_{\beta}$}
\State $\theta[\beta]^* \gets \theta[\beta]^{(k)}, \beta^* \gets \beta^{(k)}$
\EndIf
\EndFor
\State $\theta[\beta]^* \gets \theta[\beta]^{(K)}, \beta^* \gets \beta^{(K)}$
\end{algorithmic}
\end{algorithm}

\rev{
\begin{remark}[Computation efficiency of Algorithm \ref{alter_algo}]\label{remark_computation}
    Algorithm \ref{alter_algo} in essence performs coordinate-descent on the non-convex optimization problem \eqref{mod}. Doing so in general may not exhibit fast converge. Nevertheless, in our case, the number of iteration $N$ is always between 3 and 5. A typically loss curve over $\beta$ is given in Figure \ref{loss_over_beta} below. Specifically, the consecutive $\beta^{(3)}=0.76$ (after three iterations) and $\beta^{(2)}=0.78$ are close enough, so that Algorithm 2 terminates. In terms of clock time (measured on 16-inch Macbook Pro 2019), the computation per iteration is $\sim$12 seconds on the small-scale example with 36 locations and is $\sim$3.8 minutes on the large-scale example with 453 locations. Given that parameters are fixed during prediction, the proposed optimization procedure in Algorithm 2 is thus efficient. 
    
    Intuitively, we think the reason behind fast convergence is partly because the optimization problem in $\beta$ when other parameters are fixed behaves reasonably nicely---objective (8) mainly comprises of $-\log(\sum_k c_k \beta e^{-\beta t_k})+c(1-e^{-\beta t})$ for constants $c_k$ and $c$. Numerically, we often find exactly one local minimizer in reasonable range of $\beta$. 

    \begin{figure}[!b]
      \begin{center}
      \includegraphics[width=0.48\textwidth]{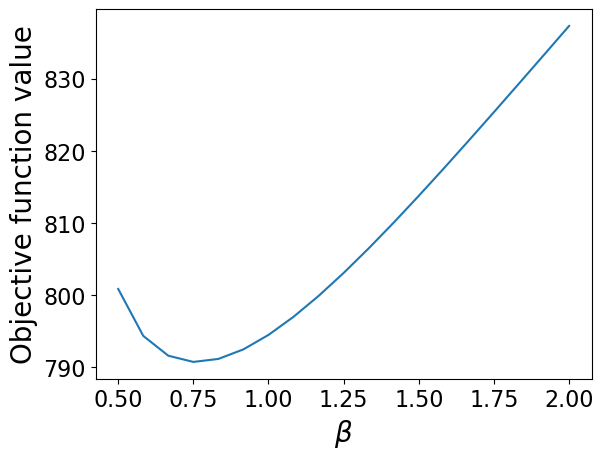}
      \end{center}
      \vspace{-0.15in}
      \caption{\rev{Objective \eqref{mod} over $\beta\in[0,2]$ on the small-scale example with $K=36$ location. The interval is discretized into 25 evenly-spaced grid points.}}
      \label{loss_over_beta}
    \end{figure}
\end{remark}
}

\subsection{Dynamic threshold selection}\label{appendix:thres}

Let $Y_{tk}\in\{1,-1\}, t\geq 1, k\in [K]$ denote the fire occurrence status in location $k$ at time $t$, where 1 indicates that a fire event occurs. Since fire incidents are rare, we also view $Y_{tk}=1$ as anomalies. Moreover, $Y_{tk}$ is fully observable after time $t$, so we have full feedback after identifying the anomalies. Inspired by the Hedging Algorithm \cite[Hegding (Algorithm 4)]{rebecca_thres}, we thus construct a dynamic threshold selection procedure in Algorithm \ref{algo:thres}, which leverages current prediction and feedback.

We explain the intuitive procedures of Algorithm \ref{algo:thres}. Overall, the algorithm updates thresholds only when the current anomaly prediction is false. It does so by increasing/decreasing the threshold if an anomaly/normal datum is estimated. Then, it projects the threshold back to a target interval determined by past predicted risks. \rev{Meanwhile, we realize in practice that due to rareness of true fire incidents and the randomness in predicted risks, there tends to be an excessive number of positive prediction, leading to a significant number of false positives. These false positives are especially undesirable and costly in the case of power system management, where power delivery facilities are mistakenly shutdown to avoid further damages. Thus, }to further control the number of false positives, we predict it as an anomaly only when the ``slope'' of increase is large enough even if a risk estimate exceeds the threshold---this procedure is highlighted in line 8: $\Delta_{tk}\geq\delta_k \text{ and } \lambda(t,k,m)>\tau_{tk}$\rev{, where $\Delta_{tk}$ is defined in \ref{eq_inc}}. We do so since true anomalies typically occur when the relative risk increase is large enough; Figure \ref{fig:loc_pred_carved} shows an example of this. The choice of $\delta_k$ may be guided by historical data (e.g., what is the lowest/largest/average rate of increase $\Delta_{tk}$ in validation data for each $k$). Furthermore, to reduce false positives, line 11 ($\tau_{tk}:=\max(\Pi(\tau_{t-1,k}+\eta_k \hat{Y}_{tk}),\lambda(t-1,k,m)/a_{1k})$) ensures that thresholds increase sufficiently quickly under sharp rise in risk estimates, even if the projection operation do not increase the risk fast enough. Lastly, line 15 (Reset $\tau_{tk}=\lambda(t,k,m)$ if $\lambda(t,k,m)\leq \lambda(t-1,k,m)/a_{2k}$)  ensures that when risk estimates drop significantly at location $k$ (e.g., under seasonal shifts from summer to fall), the algorithm resets thresholds to capture possible future rise in estimates. One can achieve different performances by tuning knobs $\{a_{1k}, a_{2k}\}$ in these two lines; in practice, larger $a_{1k}$ implies more positive anomaly estimates, and the algorithm resets thresholds less often under larger $a_{2k}$. If risk estimates are fairly constant, we recommend setting $a_{1k}, a_{2k}$ fairly close to 1. After tuning, we set other parameters as $\tau_{k,\min}=\lambda(1,k,m)/1.8,\tau_{k,\max}=\lambda(1,k,m)\times 1.8,\eta_k=(\tau_{k,\max}-\tau_{k,\min})/(T^{1.5}),\delta_k=0.05$.
\begin{algorithm}[!t]
\caption{Location-wise Dynamic Threshold Selection}
\label{algo:thres}
\begin{algorithmic}[1]
\Require{Risk estimates $\{\lambda(t,k,m)\}_{t=1}^T, \tau_{k,\min},\tau_{k,\max},\eta_k,\delta_k,a_{1k},a_{2k},$ and true anomalies $\{ Y_{tk}\}_{t=1}^T$, revealed individually after each prediction.}
\Ensure{Decision thresholds $\{\tau_{tk}\}_{t=1}^T$, anomaly estimates $\{\hat Y_{tk}\}_{t=1}^T$.}
\State Define Projection $\Pi(x):=\underset{\tau \in [\tau_{k,\min},\tau_{k,\max}]}{\arg \min} (\tau-x)^2$.
\State Initialize $\tau_{1k}=\tau_{k,\min}$ and let $\hat Y_{1k}=1$ if $\lambda(1,k,m)>\tau_{1k}$.
\If{$\hat{Y}_{1k}\neq Y_{1k}$}
\State Let $\tau_{2k}:=\max(\Pi(\tau_{1k}+\eta_k \hat{Y}_{1k},\lambda(1,k,m)/a_{1k})$
\EndIf
\For{$t=2,\ldots,T$}
\State Define increase 
\begin{equation}\label{eq_inc}
    \Delta_{tk}:=|(\lambda(t,k,m)-\lambda(t-1,k,m))/\lambda(t-1,k,m)|
\end{equation}
\If{$\Delta_{tk}\geq\delta_k \textbf{ and } \lambda(t,k,m)>\tau_{tk}$}
\State Let $\hat{Y}_{tk}=1$
\If{$\hat{Y}_{tk}\neq Y_{tk}$}
\State Let $\tau_{tk}:=\max(\Pi(\tau_{t-1,k}+\eta_k \hat{Y}_{tk}),\lambda(t-1,k,m)/a_{1k})$
\EndIf
\EndIf
\If{$\lambda(t,k,m)\leq \lambda(t-1,k,m)/a_{2k}$}
\State Reset $\tau_{tk}=\lambda(t,k,m)$.
\EndIf
\EndFor
\end{algorithmic}
\end{algorithm}

In practice, fire typically densely clusters near summer (e.g., June--August), so we also apply the following screening procedure at each $(t,k)$ before applying the algorithm. First, compute the number of fire incidents, frequency, and the gap between fire events on validation data at $k$. Second, require true statements for all three screening questions and claim no fire at $(t,k)$ if any answer is false:
\begin{enumerate}[noitemsep,topsep=0em]
    \item There had been at least one fire incident at location $k$.
    \item The number of detected fire at $k$ has not exceed the total number of fire occurred at $k$ in validation data.
    \item The time since the last positive detection is no less than the average fire occurrence gap in validation data.
\end{enumerate}
The procedures above aim to limit the number of false positives during detection based on the following observation: bumps/sudden rises in predicted risks often exist outside summer, when fire incidents rarely exist. To make better detection besides naively using an average or a sum as the metric, one may use historical data (training and validation) to predict a distribution of the total possible number of fires at $k$ in test time. Then, one can decide the total number of detection based on statistical tests over this predicted distribution. Such Bayesian-type approaches can be more systematic but may also introduce additional complication that hinders computational efficiency, so we leave it as future work.
\begin{figure}[!b]
    \begin{minipage}{0.48\textwidth}
        \centering
        \captionsetup[subfigure]{justification=centering}
        \includegraphics[width=\linewidth]{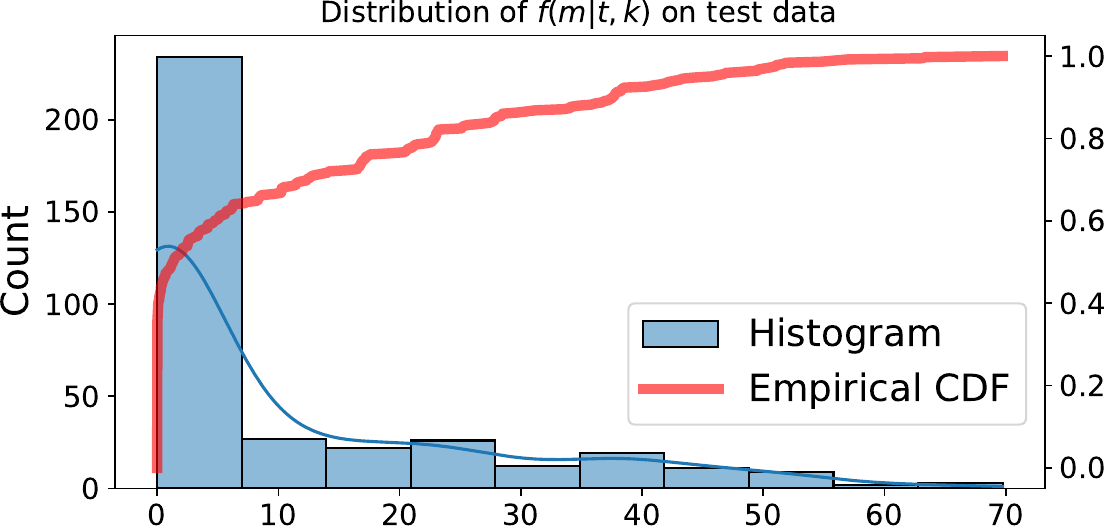}
        \subcaption{1000 Frechet random variables}
    \end{minipage}
    \begin{minipage}{0.48\textwidth}
        \centering
        \captionsetup[subfigure]{justification=centering}
        \includegraphics[width=\linewidth]{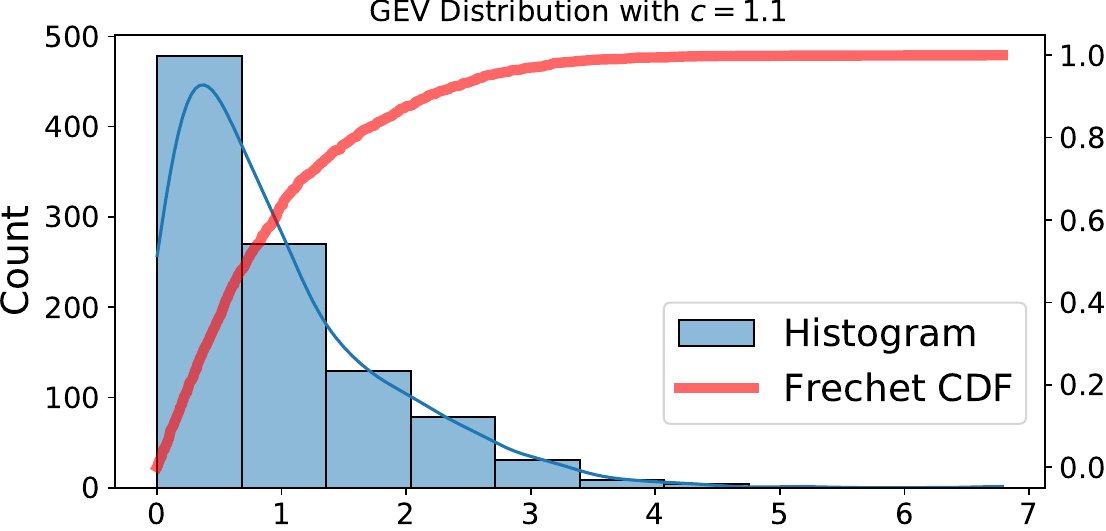}
        \subcaption{Estimated $g(m|t,k)$}
    \end{minipage}
    \caption{Compare Frechet random variables with our estimated conditional intensities at the first location of the large region in test time.}
    \label{fig:EVT}
\end{figure}

\subsection{Empirical observed connection with extreme value distribution}\label{appendix:EVT}

We observe empirically that the distribution of estimated mark influences in \NonLinearSTHawkes \ (cf. \eqref{eq:feature_extractor}) is similar to the Frechet distribution. This similarity is illustrated in Figure \ref{fig:EVT} for a Frechet distribution with the shape parameter being 1. Such a connection is useful as the Frechet distribution belongs to the family of generalized extreme value distribution (GEV), which has been used to capture the distribution of rare events, such as catastrophes \cite{sanders_2005}. \rev{In our case, fire incidents are rare events, and it is natural to expect the dependency of fire risks on marks to also follow extreme value distribution (e.g., only rare weather lead to significant impact on fire risks).} How to better incorporate such information as priors in the model belongs to future work.

\end{document}